\documentclass[a4paper,11pt,dvipsnames]{aa}  
\usepackage{graphicx}
\usepackage{txfonts}
\usepackage{natbib,twoopt}
\usepackage[breaklinks=true]{hyperref} %% to avoid \citeads line fills
\bibpunct{(}{)}{;}{a}{}{,}             %% natbib format for A&A and ApJ
\makeatletter
  \newcommandtwoopt{\citeads}[3][][]{\href{http://adsabs.harvard.edu/abs/#3}%
    {\def\hyper@linkstart##1##2{}%
     \let\hyper@linkend\@empty\citealp[#1][#2]{#3}}}
  \newcommandtwoopt{\citepads}[3][][]{\href{http://adsabs.harvard.edu/abs/#3}%
    {\def\hyper@linkstart##1##2{}%
     \let\hyper@linkend\@empty\citep[#1][#2]{#3}}}
  \newcommandtwoopt{\citetads}[3][][]{\href{http://adsabs.harvard.edu/abs/#3}%
    {\def\hyper@linkstart##1##2{}%
     \let\hyper@linkend\@empty\citet[#1][#2]{#3}}}
  \newcommandtwoopt{\citeyearads}[3][][]%
    {\href{http://adsabs.harvard.edu/abs/#3}
    {\def\hyper@linkstart##1##2{}%
     \let\hyper@linkend\@empty\citeyear[#1][#2]{#3}}}
\makeatother

\usepackage[T1]{fontenc}
\usepackage[utf8]{inputenc}
\usepackage{color}
\usepackage{xcolor}
\usepackage[mathscr]{euscript}

\hypersetup{linkcolor=red,citecolor=red,filecolor=cyan,urlcolor=magenta}

\usepackage{soul}

\begin{document}

   \title{ISPY - NaCo Imaging Survey for Planets around Young stars}
  \subtitle{CenteR: the impact of centering and frame selection}

   \author{N.~Godoy
          \inst{1,2}
          \and
          J.~Olofsson
          \inst{1,2}
          \and
          A.~Bayo
          \inst{1,2}
          \and
          A.~C.~Cheetham
          \inst{3,4}
          \and
          R.~Launhardt
          \inst{4}
          \and
          G.~Chauvin
          \inst{5}
          \and
          G.~M.~Kennedy
          \inst{6,7}
          \and
          S.~S.~Brems
          \inst{8}
          \and
          G.~Cugno
          \inst{9}
          \and
          J.~H.~Girard
          \inst{13}
          \and
          Th.~Henning
          \inst{3}
          \and
          A.~Müller
          \inst{3}
          \and
          A.~Musso~Barcucci
          \inst{3}
          \and
          F.~Pepe
          \inst{4}
          \and
          S.~P.~Quanz
          \inst{9}
          \and
          A.~Quirrenbach
          \inst{8}
          \and
          S.~Reffert
          \inst{8}
          \and
          E.~L.~Rickman
          \inst{4,10}
          \and
          M.~Samland
          \inst{3,12} 
          \and
          D.~Ségransan
          \inst{4}
          \and
          T.~Stolker
          \inst{9}
     }

    \institute{
         Instituto de F\'isica y Astronom\'ia, Facultad de Ciencias, Universidad de Valpara\'iso, Av. Gran Bretaña 1111, Playa Ancha, Valpara\'iso, Chile \\ %1
         \email{nicolas.godoy@postgrado.uv.cl}
         \and
         N\'ucleo Milenio de Formaci\'on Planetaria – NPF, Universidad de Valpara\'iso, Av. Gran Bretaña 1111, Valpara\'iso, Chile  %2
         \and
         Max-Planck-Institut für Astronomie, Königstuhl 17, 69117 Heidelberg, Germany  %3
         \and
         Observatoire Astronomique de l’Université de Genève, 51 Ch. des Maillettes, 1290 Versoix, Switzerland %4
         \and
         Université Grenoble-Alpes, CNRS, IPAG, F-38000 Grenoble, France %5
         \and
         Department of Physics, University of Warwick, Coventry, CV4 7AL, UK %6
         \and
         Centre for Exoplanets and Habitability, University of Warwick, Gibbet Hill Road, Coventry CV4 7AL, UK %7
         \and
         Landessternwarte, Zentrum für Astronomie der Universität Heidelberg, Königstuhl 12, 69117 Heidelberg, Germany %8
         \and
         Leiden Observatory, Leiden University, Niels Bohrweg 2, 2333 CA Leiden, The Netherlands %9
         \and
         European Space Agency (ESA), ESA Office, Space Telescope Science Institute, 3700 San Martin Drive, Baltimore, MD 21218, USA %10
         \and
         Department of Astronomy, Stockholm University, Stockholm, Sweden %12
         \and
         Space Telescope Science Institute, 3700 San Martin Drive, Baltimore, MD 21218, USA %13
}

   \date{Accepted for publication in A\&A, November 2021.}

  \abstract
  % context heading
{Direct imaging has made significant progress over the past decade, in part thanks to new generation instruments and excellent adaptive optic systems, but also thanks to advanced post-processing techniques. The combination of both allowed the detection of several giant planets with separations as close as $0.2$ arcsec with contrast typically reaching $9-10$ magnitudes at near-infrared wavelengths. Observing strategies and data-rate vary from instrument to instrument and wavelength, with $L$ and $M$ band observations yielding tens of thousands of images to be combined.}
  %Aims
{We present a new approach, tailored for VLT/NaCo observations performed with the Annular Groove Phase Mask - AGPM - coronagraph, but that can be applied to other instruments using similar coronagraphs. Our pipeline aims at improving the post-processing of the observations on two fronts: identifying the location of the star behind the AGPM to better align the science frames as well as performing frame selection.}
  %Method
{Our method relies on finding the position of the AGPM in the sky frame observations, and correlate it with the circular aperture of the coronagraphic mask. This relationship allows us to retrieve the location of the AGPM in the science frames. We are then able to model the ``torus'' shape visible in the sky-subtracted science frames, as a combination of negative and positive 2D Gaussian function. The model provides additional information that is useful to design our frame selection criteria.}
  % results heading 
{We tested our pipeline on three targets ($\beta$\,Pictoris, R\,CrA, and HD\,34282), two of them having companions at intermediate and close separations, and the third one hosting a bright circumstellar disk. We find that the centering of the science frames has a significant impact on the signal to noise ratio (S/N) of the companions. Our results suggest that the best reduction is achieved when performing the principal component analysis centered on the location of the AGPM and de-rotating the frames centered at the location of the star before collapsing the final datacube. We improve the S/N of companions around $\beta$\,Pictoris and R\,CrA by $24\pm3$\,\% and $117\pm11$\,\% respectively, compared to other state-of-the-art reductions. We find that the companion position for all the centering strategies are consistent within $3\sigma$. Finally, we find that even for NaCo observations with tens of thousands of frames, frame selection yields just marginal improvement for point sources but may improve the final images for objects with extended emission such as disks.}
  % conclusions 
{We propose a novel approach to identify the location of the star behind a coronagraph even when it cannot easily be determined by other methods. We led a thorough study on the importance of frame selection, concluding that the improvements are marginal in most cases but may yield better contrast in some specific cases. Our approach can be applied to the wealth of archival NaCo data, and, assuming that the field of view includes the edges of the coronagraphic mask, its implementation can be adapted to other instruments with similar coronagraphs as the AGPM used on NaCo (e.g., Keck/NIRC2, LBT/LMIRCam).}


   \maketitle
%
%-------------------------------------------------------------------

\section{Introduction}

Thousands of exoplanets have been discovered in the past decades using a variety of observational techniques. Directly measuring the brightness of the planets using high-contrast imaging techniques, provides crucial information on their temperatures and chemical compositions, and therefore on their possible formation history. This is particularly achievable when using spectro-photometry in the near- and mid-infrared (e.g., \citealt{Samland+2017}). In addition, it allows to probe an unique parameter space, in particular at large orbital separations (typically >10 au for nearby stars). Several dedicated surveys aim at finding young giant gaseous planets using the direct imaging technique (see \citealt{Bowler_2016} for an overview of previous large imaging surveys). Such surveys take advantage of the fact that gaseous giant planets at an early age ($<100$\,Myr) are bright at infrared wavelengths because of the energy released during the contraction of their atmospheric envelope (see, for instance, \citealt{Marley+2007}). For this reason, many of the recent, present and future direct imaging instruments focus on near- and mid- infrared observations (for example SCExAO, \citealt{Jovanovic+2015}; MagAO-X, \citealt{Males+2018}; SPHERE \citealt{Beuzit+2019}; GPI, \citealt{Macintosh+2014}). In particular, the ``NaCo Imaging Survey for Planets around Young stars'' (NaCo-ISPY, \citealt{ISPY+2020}) is a survey aiming at finding new young ($<100$\,Myr) sub-stellar ($\lesssim72M_{\rm Jup}$) companions at $L^{\prime}$ filter ($\sim3.8\,\mu$m) using the NaCo instrument (NAOS+CONICA, \citealt{Rousset+2003},\citealt{Lenzen+2003}) at the Very Large Telescope (VLT), Paranal Observatory, Chile.

The main challenge to detect and characterize faint companions around bright stars is to reach very large contrast values in the immediate vicinity of the star. Using a coronagraph to block the stellar light has proven to be a reliable strategy, improving the contrast by several orders of magnitude (e.g. \citealt{Chauvin_2018}), as close as a few $\lambda$/D from the star (typically $>$100\,mas with current planet imagers). In addition, as mentioned in \citet{ISPY+2020}, the contrast between a planet and its host star is more favorable at mid-infrared wavelengths. However, the observations are still largely dominated by the stellar halo and the point spread function (PSF), obscuring the planet signal in the images, so several observational techniques were designed to tackle this challenge: angular differential imaging (ADI, \citealt{Marois+2006}), reference differential imaging (RDI, \citealt{Smith_and_Terrile_84}), and spectral differential imaging (SDI, \citealt{Marois+2000}, \citealt{Sparks_and_Ford_2002}). Since ISPY uses only one filter and has no spectral information, it uses ADI as it allows the temporal subtraction of the PSF and the quasi-static speckles using the data of the objects themselves as reference. Since the sky rotates during the observing sequence, while the PSF and the speckles remain quasi-static, subtracting the median from the data cube essentially removes their contribution in the final image (classical ADI, \citealt{Marois+2006}). The suppression of the stellar PSF is significantly improved by using advanced post-processing algorithms and, for the forward-modelling approaches, prior information on the noise statistics and the ADI planetary signal itself. Some of these algorithms are: the locally optimized combination of images (LOCI, \citealt{Lafreniere+2007}), the Karhunen-Lo\'eve image projection or principal component analysis (KLIP, \citealt{Soummer+2012}; PCA, \citealt{Amara_and_Quanz_2012}; KLIP-forward modelling or KLIP-FM, \citealt{Pueyo_2016}), the maximum likelihood-based ANDROMEDA (\citealt{Mugnier+2009}; \citealt{Cantalloube_Mouillet_Mugnier+2015}), the local low-rank, sparse, and Gaussian noise component decomposition (LLSG, \citealt{Gomez-Gonzalez+2016}), a dedicated PSF reference library construction (e.g. \citealt{Xuan_2018}, \citealt{Ruane_2019} or \citealt{Bohn_2019}), to perform reference differential imaging (RDI, \citealt{Lafreniere_2009}, \citealt{Soummer_2011}, \citealt{Gerard_Marois_2016}), the exoplanet detection based on patch covariances (PACO, \citealt{Flasseur+2018}, \citealt{Flasseur+2020}), and more recently the temporal causal regression model (lightcurve) approach (TRAP, \citealt{Samland+2021}).

Surveys with very large samples allow us to put statistically robust constraints on the occurrence of giant planets around young stars, and therefore provide critical inputs to inform and refine planet formation models (see, for example, \citealt{SHINE+2020}). As many bright ($L<$6\,mag) stars were included in the ISPY sample, it was possible to observe a large fraction of them ($\sim70$\%) using the annular groove phase mask (AGPM) vector vortex coronagraph (\citealt{Mawet+2013a}, \citealt{Absil+2014}). The position of the star is a crucial parameter in the post-processing of ADI observations since it is the reference center to align and combine all the images. However, as soon as the star is placed behind the AGPM, its exact position can no longer be measured accurately, and this represents a challenge for high-resolution imaging observations, especially for a first generation direct imaging instruments such as NaCo. Second-generation instruments tackle this problem using the deformable mirror to produce diffractive attenuated copies of the stellar PSF, called satellite spots (SPHERE, \citealt{Beuzit+2019}; SCExAO/CHARIS, \citealt{Groff+2015}), or using diffracted images of the star produced by a reference grid inside the optical path of the instrument (GPI, \citealt{Macintosh+2014}), to determine the position of the star behind the coronagraph with sub-pixel precision (reaching $0.1-0.2$ pixels in the case of SPHERE in H-band\footnote{According with the appendix A.12, page 116 in the VLT/SPHERE User Manual Issue: 106}).

Ground-based observations at mid-infrared wavelengths are performed with a very short detector integration time (DIT) to avoid saturating the detector, because of the significant sky thermal emission. The typical on-source time per target for the ISPY observing strategy is between $2$ to $4$ hours, leading to $10,000$ to $30,000$ frames per target. Because the observing conditions vary throughout the night, or the coronagraph is slightly drifting (see Section\,\ref{sec:Cent}), the quality of those thousands of frames will significantly vary. It is therefore relevant to assess whether removing the worst frames, while keeping the most homogeneous ones, can improve the final reduction or not. Different frame selection techniques have been proposed to try and improve the contrast or increase the signal-to-noise of any point source detection at different wavelengths. For example, \citet{Stolker+2019} used a selection based on the sky brightness, \citet{Keppler+2018} used the PSF variation of the stars in the field of view, while \citet{Bohn+2020} discarded images with poor AO correction. There are therefore several approaches to perform some sort of frame selection, but overall those approaches are usually on the conservative side, and there has not been an in-depth study of the potential benefits of performing more aggressive frame selection on large data cubes.

In this paper, we present a new technique aiming to accurately find the position of the star behind the coronagraph for the particular case of NaCo observations with the AGPM. A good determination of the position of the star is essential for astrometric studies and a better extraction of the flux of potential companions. We also present our approach to register the quality of the science frames, and study the impact of performing frame selection on the final reduction.

The paper is structured as follows: Section\,\ref{sec:O-DR} summarizes the observation strategy and data reduction. In Section\,\ref{sec:pipeline} we explain the new centering technique and the determination of the position of the star behind the coronagraph. We also explain our approach to register and select the science frames. In Section\,\ref{sec:C-test}, we benchmark our centering technique using observations of a star with a known companion. We apply our pipeline to two known companions ($\beta$\,Pictoris\,b and R\,CrA\,b) and present in Section\,\ref{sec:Appl} the improvements on the signal-to-noise ratio of those point sources. In Section\,\ref{sec:concl} we summarize our findings and conclusions on the importance of centering and frame selection for ADI observations.

\section{Observations and data reduction} \label{sec:O-DR}

In this section, we describe the observing strategy, the standard (cosmetic) reduction steps, and the post-processing to obtain the final images. In addition, we describe the effective shape of the coronagraphic image in the corrected science data, that will be used later on for the centering algorithm and registration of frames.

\subsection{Observational strategy}

The stars HD\,34282, R\,CrA, and HD\,104237 were observed as part of the NaCo-ISPY program \citep{ISPY+2020}, during the following nights: 2016-11-07, 2017-05-19, and 2017-05-16. The observations were performed under different weather conditions varying from good to bad seeing (ranging between $0.34\arcsec$ and $1.54\arcsec$) as summarized in Table\,\ref{Summary_table}. We used the L27 camera (pixel scale $\sim27.2$\,mas\,pix$^{-1}$) with the $L^{\prime}$ filter ($\lambda_{c}=3.8\,\mu$m, $\Delta\lambda=0.62\,\mu$m) and the AGPM coronagraph for all three stars. The observing strategy was already described in \citet{ISPY+2020}, but as a brief summary, it consists in obtaining multiple datacubes of $100$ images each with the star manually placed behind the AGPM (hereafter, science sequence). At the end of each science sequence the field of view is shifted so that it does not include the star (hereafter, sky sequence). The purpose of the sky sequence is to estimate the thermal background and subtract it from the science observations. Each science-sky sequence lasts between $3-5$ minutes and is repeated tens of times for each star. At the beginning and end of every observation, we also obtained a separate set of images with the star placed at different regions of the detector (far from the AGPM), with shorter exposure times to measure the stellar PSF for photometric calibration (hereafter, photometry frames or images). Since the third quadrant (bottom left corner of the camera) contains persistent bad columns, it is not used for the photometry observations.

$\beta$\,Pictoris was observed on the 2013-02-01 (observations not part of the ISPY program), with the AGPM, under rather poor conditions. The data were presented in \cite{Absil+2013} and re-reduced in \cite{Stolker+2019}, using the same strategy as the one used within ISPY, but for those observations, the DIT was set to $0.3$\,s and the number of exposures per science cube was set to $200$. The flux calibration strategy was also slightly different: in ISPY, the star is moved to the three good quadrants of the detector, but for those observations, sky measurements were taken in-between photometry frames. The observations were executed using a window size of $768\times770$ pixels, that, in combination with the DIT, lead to a loss of $\sim10\%$ of the frames. The total number of science sequences, the exposure times and the weather conditions are summarized in Table \ref{Summary_table} for all stars.

\begin{table*}
\caption{Log of the observations.}   
\centering                        
\begin{tabular}{l c c c c c}        
\hline\hline                
  & HD\,34282 & R\,CrA & HD\,104237 & $\beta$\,Pictoris  \\     
\hline                        
 Obs. date & 2016-11-07 & 2017-05-19 & 2017-05-16 & 2013-02-01 \\
 Wind Speed\tablefootmark{a} [m/s] & $2.15$-$5.18$\,($4.00$) & $5.03$-$7.80$\,($6.09$) & $12.98$-$17.08$\,($15.43$) & $0.53$-$4.28$\,($2.28$) \\
 Coherence time\tablefootmark{a} [ms] & $4.8$-$17.3$\,($10.7$) & $2.9$-$5.8$\,($4.2$) & $1.8$-$3.7$\,($2.3$) & $1.03$-$2.43$\,($1.76$) \\
 Seeing\tablefootmark{b} [\arcsec]  & $0.69$-$1.29$\,($0.87$) & $0.67$-$0.92$\,($0.77$) & $0.97$-$1.43$\,($1.21$) & $0.84$-$1.57$\,($1.13$) \\
 Field Rotation [deg] & 118.26 & 35.86 & 45.18 & 82.97 \\
 DIT\tablefootmark{c} [sec] & 0.25 & 0.35 & 0.35 & 0.2 \\
 NDIT & 100 & 100 & 100 & 200 \\
 Total exposure time [min.] & $\sim135$ & $\sim70$ & $\sim152$ & $\sim114$ \\
 Total number of Frames & 32500 & 12100 & 26100 & 34277 \\
 Program ID & 198.C-0612(A) & 199.C-0065(A) & 199.C-0065(A2) & 60.A-9800(J) \\
\hline                                   
\end{tabular}
\tablefoot{
For the wind speed, coherence time, and seeing, the minimum and maximum value registered are tabulated, while the median registered value is between parenthesis.
 \tablefoottext{a}{DIMM measured values during the observations.}
 \tablefoottext{b}{Seeing measured on the image analysis detector (at the pointing location).}
 \tablefoottext{c}{DIT\,=\,Detector Integration Time.}
}
\label{Summary_table}%
\end{table*}

\subsection{Cosmetic corrections}\label{subsec:cosmetics}

To prepare the data for the analysis, it is necessary to perform cosmetic corrections. First, we corrected the images for the dark current subtracting the suitable master darks. Master darks were produced by calculating the median of the dark frames grouped by the different exposure times used in an observing sequence (photometry, science, sky, flat field). Every image is corrected by the corresponding master dark by subtracting it from the image. The second correction corresponds to normalizing the response of all pixels. To that end, we construct a master flat field by median averaging all the sky flat images (flat images are usually taken once per observing run and correspond to sky flats), and dividing by the mean value of counts. The correction for the flat field is done simply by dividing every image by their respective master flat field. At the end, all science, sky and photometry images were corrected by master dark and master flat field. We also create a bad pixel map using the master dark and performing $\sigma$-clipping to select the outlier pixels (with $\sigma=5$). The bad pixels are then corrected for by using a bicubic interpolation for every image. In the end, the sky mean is computed for every sky datacube using all available images. Then, the science images are sky-subtracted using the sky observations that are closest in time. We note that performing the sky subtraction this way is possible because of the negligible motion of the circular aperture and of the AGPM between consecutive science-sky sequences (see Appendix\,\ref{Apx:CA}).

\subsection{The effective shape of the coronagraphic image}\label{sec:shape}

When subtracting the sky background from the science images, we are also removing the thermal contribution of the AGPM (see \citealp{Absil+2016}) and dust/debris located in the optical path of the telescope and instrument. A bright torus becomes visible around the coronagraphically rejected central star image. This torus is the result of the finite bandwidth of the AO, both spatial and temporal. Wavefront aberrations outside this bandwidth are not corrected for and lead to off-axis modes that are not rejected by the AGPM. Although the AGPM efficiently cancels out the on-axis light from the star, its transmission rapidly increases for off-axis light (see Fig. B.1 in \citealp{ISPY+2020}). The shape of the torus strongly depends on the alignment between the star and the AGPM, the performance of the adaptive optics system, and in general, the weather conditions. Both the AGPM and the star can move during an observing sequence (see Section\,\ref{sec:pipeline}), with the motion of the AGPM being negligible for consecutive science-sky frames (see Fig.\,\ref{fig:CA_vs_time} and Fig.\,\ref{fig:AGPM-CA_vs_time} in the Appendixes\,\ref{Apx:CA} and\,\ref{Apx:AGPM-CA}, respectively). For this reason, it is important to: (1) find the center position of the AGPM on the detector and, (2) find the star position with respect to the AGPM center.

\subsection{Principal Component Analysis}\label{PCA-sec}

The PCA and the de-rotation processes are performed using the \texttt{PynPoint}\footnote{PynPoint version 0.6.2: \url{https://pynpoint.readthedocs.io/en/latest/}} package (\citealp{Amara_and_Quanz_2012}; \citealp{Stolker+2019}). The PCA process consists of creating an orthonormal basis of images (or eigen-images) representative of the ensemble of frames. Ultimately, the goal is to build an accurate representation of the stellar PSF and quasi-static speckles, which will be subtracted from each individual frame. Ideally, due to the sky rotation during the pupil-stabilized observations providing angular diversity with respect to the speckle noise, any real astrophysical signal should survive the process. Once the estimated PSF and quasi-static speckles have been removed, each frame is derotated according to their parallactic angle and all the frames are median-combined to recover the signal in the field of view (for more details, see \citealp{Amara_and_Quanz_2012} and \citealp{H-Q-A-M+18}). For the method to work, there is a compromise between the total field rotation (effective rotational angle) and the minimization of the flux loss of the astrophysical signal. The final images are obtained using a specific number of principal components. Given that the computational time necessary for the entire process is proportional to the size of the datacube, we performed our tests using different image sizes, between $39 \times 39$ and $121 \times 121$ pixels. For a given star, several images are produced, with different numbers of principal components to find the maximum S/N of the point sources in the field of view (if any) and to study the impact of the frame selection process as a function of the number of principal components. Their number varies in the range $1$ to $30$ or $1$ to $150$ (depending on the size and total number of frames).

\section{CenteR algorithm}\label{sec:pipeline}

In this section, we introduce a new algorithm, CenteR\footnote{Available at \url{https://github.com/Nico-Godoy/CenteR.git}}, designed to try to improve the data reduction process for angular differential imaging observations, focusing on observations taken with a coronagraph (AGPM). Our approach is two-fold; first we aim at better estimating where the star lies behind the AGPM; and second, we perform frame registration to characterize the quality of the science frames to later on perform frame selection.

\subsection{Calibrating the AGPM detector location}\label{sec:Cent}

One critical aspect when performing high-contrast imaging with ADI, is to accurately determine the position of the star behind the coronagraph. This is especially a challenging problem for first-generation instruments such as NaCo, since \cite{Mawet+2013a}, and \cite{Huby+15} showed that the location of the AGPM is not always the same as the one registered on the detector, on top of a less stable and dynamic adaptive optics correction. We therefore present here a new method to determine the position of the star behind the AGPM. When the star is aligned behind the AGPM, the resulting shape is similar to a torus for sky-subtracted science images (see previous Section\,\ref{sec:shape} and Figure\,\ref{fig:Real_vs_Fit}). Therefore, the most straightforward way to obtain the position of the star is to fit a positive and a negative 2D Gaussian profile, modeling the aforementioned torus profile. However, the problem is highly degenerate if all the parameters are free to vary (around $13$, depending on the assumptions), leading to poor constraints on the position of the star (uncertainty of the order of $0.65$\,pixels when considering all the parameters as free). We present a novel approach that helps determine the position of the star, by fixing some of the free parameters, hence alleviating the degeneracies of the modeling. In particular, we fix some parameters of the negative 2D Gaussian profile, related to the AGPM.

Our approach consists of finding the position of the AGPM independently, to later on keep it fixed during the rest of the modeling of the torus. In the sky images, the AGPM is bright (see Fig.\,\ref{fig:SkyImage}), due to its thermal emission, and as noted by \citet{Absil+2016}, it directly corresponds to the vortex center, which is not contaminated by the contribution of the star. By fitting a positive 2D Gaussian profile, for example, it is possible to accurately determine its position. However, for the science frames, the star is behind the AGPM, making it more difficult to determine the position of the AGPM. In Figure\,\ref{fig:SkyImage}, the AGPM is clearly visible close to the center, along with the borders of a circle that correspond to the edges of the $15$\arcsec circular aperture of the coronagraphic mask. We can correlate the center of this circular aperture with the position of the AGPM in the sky images. It is then possible to find the position of the AGPM in the science images by measuring the center of this circular aperture, following the next steps:

\begin{enumerate}
 \item Measure the AGPM position in the sky frames.
 \item Measure the center of the circular aperture in the sky frames.
 \item Relate both centers to obtain the transformation between the position of the AGPM with respect to the center of the circular aperture.
 \item Measure the center of the circular aperture in the science frames without correcting by the thermal sky emission.
 \item Use the transformation found in (3) to obtain the AGPM position in the science frames.
\end{enumerate}

\begin{figure*}
\centering
\includegraphics[width=8.5cm]{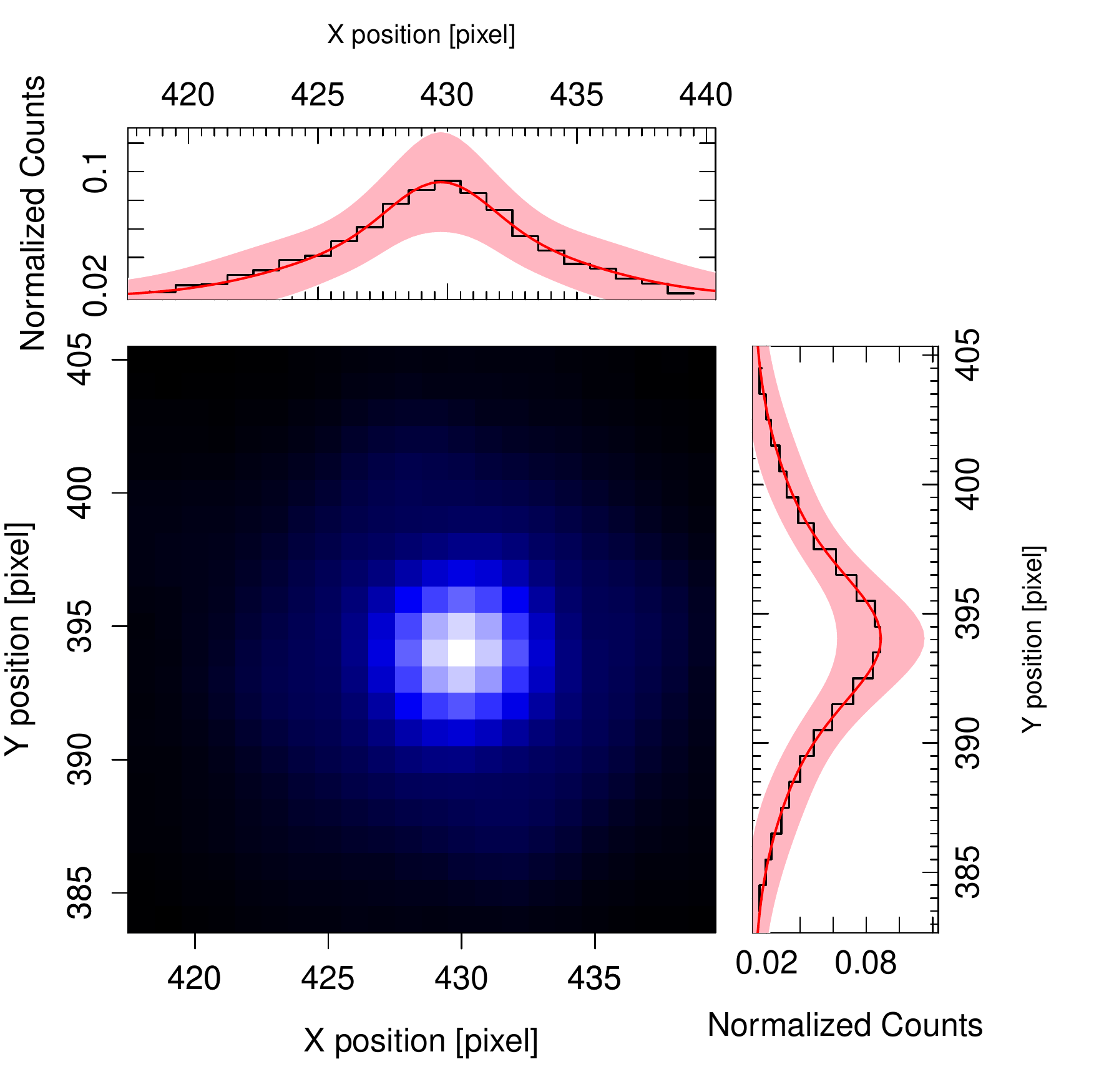}
\includegraphics[width=8.5cm]{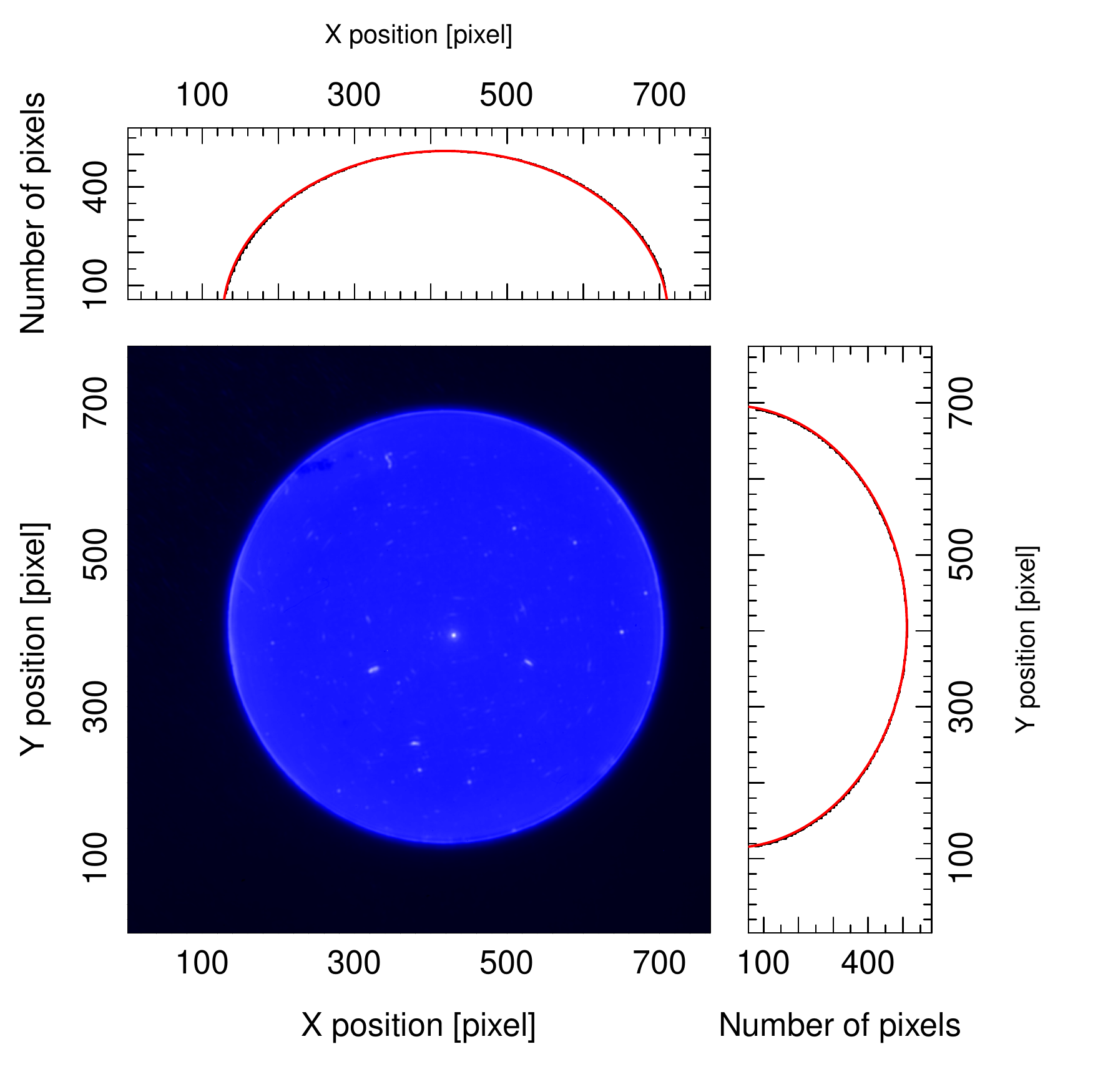}
\caption{\textit{Left}: Modeling of the thermal emission profile of the AGPM along the X- and Y-axis to obtain the coordinate center for a cleaned sky image. \textit{Right}: Modeling of the circular aperture along the X- and Y-axis to estimate the coordinates of its center in the same image. The colored area corresponds to a $95$\% confidence interval.}
\label{fig:Fitting_BC-AGPM_cnter}%
\end{figure*}

We use the sky images to estimate the relationship between the position of the AGPM with respect to the position of the center of the circular aperture. Using the collapsed frames helps increase the signal of both the aperture and AGPM, therefore minimizing the final uncertainty. This can only be achieved because of the lack of significant motion of both positions in one sequence (for ISPY, $100$ frames per sequence). The AGPM position is fitted by collapsing the image along the X and Y axis to obtain a more robust solution for the position, increasing the signal of the AGPM and making the modeling faster (see Fig.\,\ref{fig:Fitting_BC-AGPM_cnter}). 
However, 2D fitting is also possible, and depending on the model used and the number of free parameters, uncertainties of the order of $0.07$ to $0.15$\,pixels can be reached, while our approach yields uncertainties of $0.1-0.15$\,pixels. In addition, we find that both the 1D and 2D models are consistent. The solution using the 1D model is obtained by fitting two normal distributions using a non-linear least square method, minimizing the following equation:

\begin{equation}
    \eta = \sum_{i=1}^{M} \left( Z_\mathrm{i} - I_{1}\mathrm{exp}\left[-\xi_{1}\right]  - I_{2}\mathrm{exp}\left[-\xi_{2}\right] - Z_\mathrm{Bkg}   \right)^2.
\end{equation}\label{eq:AGPM-center fitting}

The value $Z_\mathrm{i}$ corresponds to the normalized sum of counts per column (or row), $\xi_{1}=(X_\mathrm{i}-\mu_\mathrm{AGPM})^2/2\sigma_{1}^2$, $\xi_{2}=(X_\mathrm{i}-\mu_\mathrm{AGPM})^2/2\sigma_{2}^2$, and $X_\mathrm{i}$ the position of each column (or row). $\mu_\mathrm{AGPM}$ corresponds to the position of the AGPM, $\sigma_{1}$ and $\sigma_{2}$ the standard deviation of both Gaussian, $I_{1}$ and $I_{2}$ the scaling factor for both profiles, and $Z_\mathrm{Bkg}$ is the background level, all of them free parameters. The variable $M$ corresponds to the number of columns (or rows). The modeling uses two Gaussian to account for the thermal emission of the AGPM and its halo (see Fig.\,\ref{fig:Fitting_BC-AGPM_cnter}). The results show that the movement of the AGPM during a normal sky sequence of $100$ images remains negligible.

To estimate the position of the circular aperture, we first identify the pixels that reside inside the aperture. To that end, we compute both the median absolute deviation ($\sigma_\mathrm{image}$) and the median using the entire image, and then for each column along the X-axis $X_i$ (or row along the Y-axis, respectively), we count the total number of pixels $N_\mathrm{i}$ that are three times $\sigma_\mathrm{image}$ above the median\footnote{Or $10$ times $\sigma_\mathrm{image}$ below the median when the observations are performed with windowing. In that case, there are much fewer pixels outside of the aperture, and therefore the median no longer is representative of the background level.}. If we were to use a linear sampling along the X-axis, the wings of the distribution would be under-sampled, which would then bias the fitting results. The sampling therefore follows a cosine distribution to more evenly distribute the number of points in polar coordinates. To find the coordinate $X_\mathrm{CA}$ and radius $R_\mathrm{CA}$ of the aperture, we then minimize the following quantity:

\begin{equation}
    D = \sum_{i=1}^N \sqrt{ (X_i - R_\mathrm{CA} \cos(\alpha_{i}) - X_\mathrm{CA})^2 + (N_i / S_\mathrm{CA} - R_\mathrm{CA} \sin(\alpha_{i}))^2,}
\end{equation}\label{eq:CA-fitting_model}

where $S_\mathrm{CA}$ is a scaling factor to account for slightly elongated shapes (i.e the aperture is not perfectly circular), and $\alpha_{i} = \arctan[(X_i-X_\mathrm{CA}) S_\mathrm{CA} / N_\mathrm{i}]$. This transformation allows us to work in polar coordinates from the center of the aperture, and the fitting is performed for the X- and Y-axis independently. Figure\,\ref{fig:Fitting_BC-AGPM_cnter} shows examples of fitting the AGPM position in a sky frame (left panel) and of the circular aperture (right panel).

\begin{figure}
\centering
\includegraphics[width=8cm]{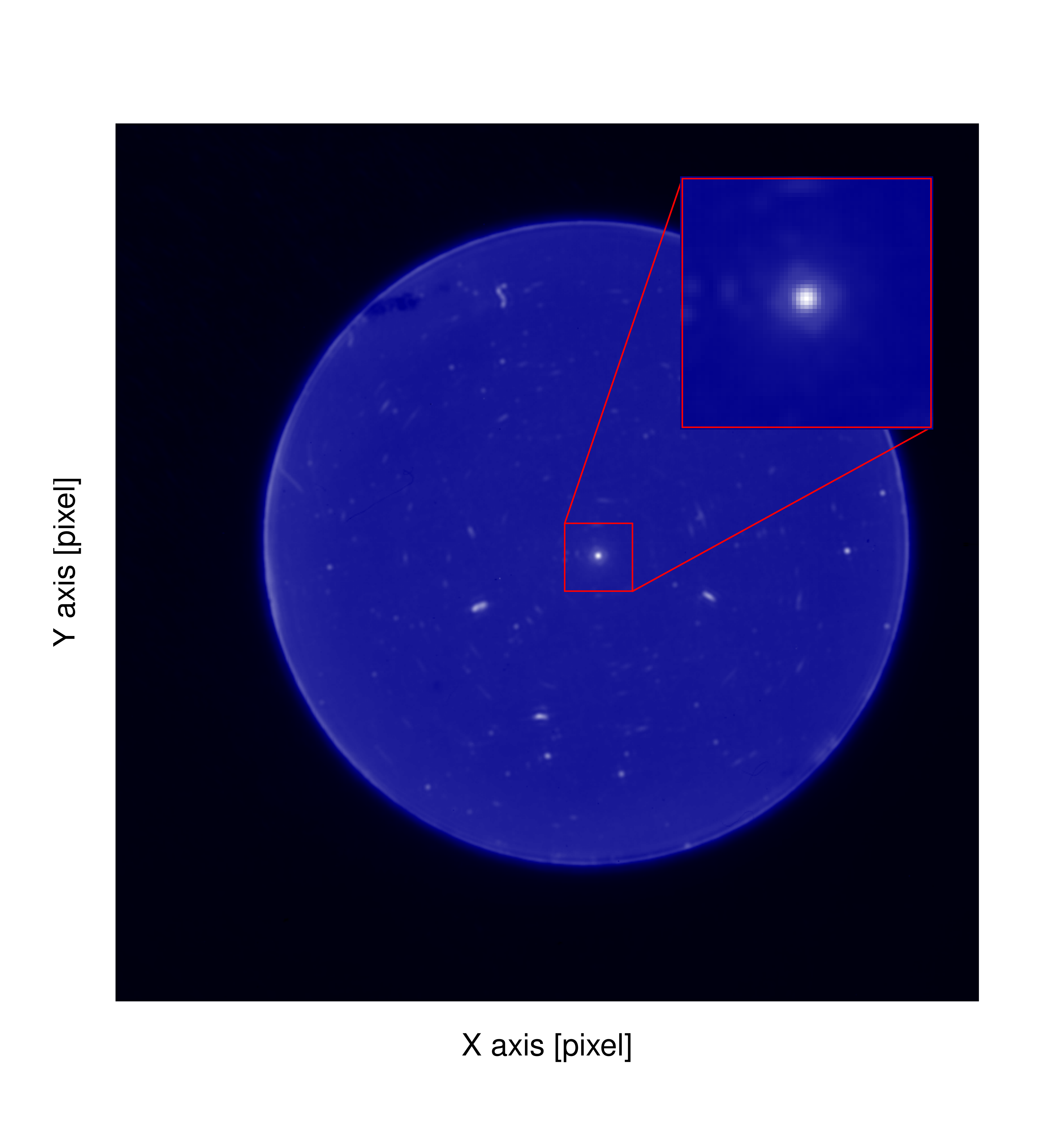}
\caption{An example of a sky image showing the circular aperture in blue (observing date 08-11-2016). The top-right inset shows a zoom of the central part, with the AGPM thermal emission in the center. The ``white elongated spots'' corresponds to dust or dirt in the optical path of the instrument or on the camera, and are emitting at near-IR wavelengths.}
\end{figure}\label{fig:SkyImage}%

\begin{figure*}
\centering
\includegraphics[width=16.0cm]{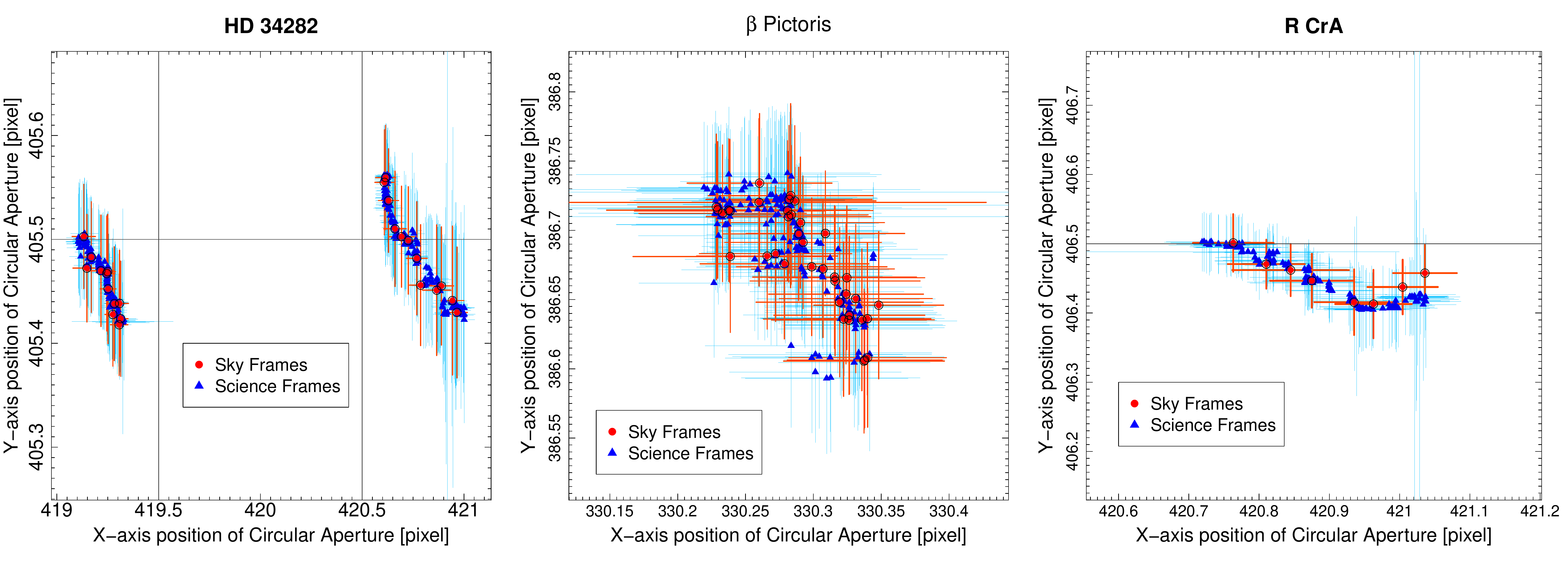}
\caption{Position of the center of the circular aperture (with respect to the camera), during the observation sequence for sky and science (red and blue, respectively) images, for HD\,34282 on the left, $\beta$\,Pictoris on the middle and R\,CrA on the right. The axes scale of the left, middle and right panels are 1:5, 1:1 and 1:3, respectively. The grid represent pixels, and the colored crosses correspond to the uncertainty obtained in the fit.}
\label{fig:BC-Sc-Sk}%
\end{figure*}

Figure\,\ref{fig:BC-Sc-Sk} shows the position of the center of the circular aperture for science and sky images (blue triangles and red circles, respectively) for the observations of HD\,34282 and R\,CrA (left and right, respectively). For HD\,34282, the observing sequence was interrupted, which explains the two separate blocks of points, indicating that the coronagraphic mask actually moves between observing sequences. The Figure shows that the center of the circular aperture for both the science and sky frames follows the same path, but also that the center of the circular aperture slightly shifts during the observing sequence. 

\begin{figure*}
\centering
\includegraphics[width=16.0cm]{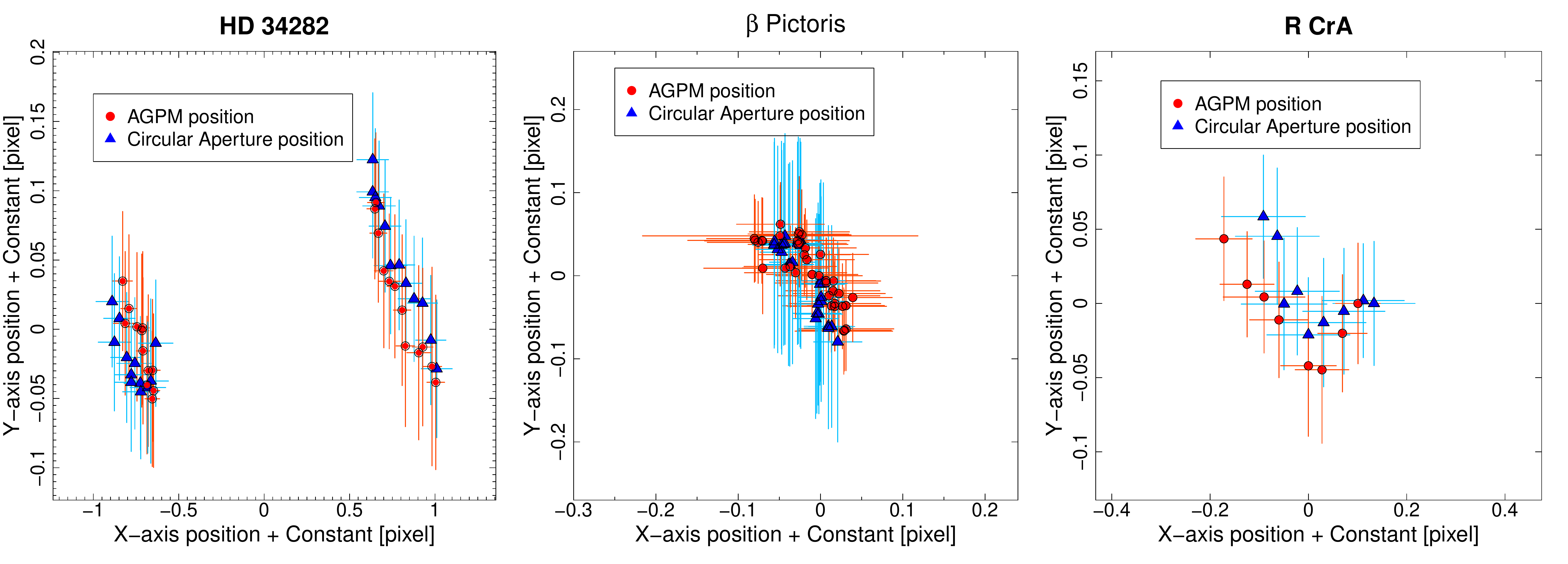}
\caption{\textit{Left:} position of the AGPM and circular aperture with respect to their median position for sky frames only, for HD\,34282. \textit{Middle:} same for $\beta$\,Pictoris. \textit{Right:} same for R\,CrA. The axes scale of the left, middle and right panels are 1:8, 1:1 and 1:3, respectively. The colored crosses correspond to the uncertainty obtained in the fit.}
\label{fig:AGPM-BC_pos}%
\end{figure*}

For the sky frames only, figure\,\ref{fig:AGPM-BC_pos} shows the position of the AGPM and the center of the circular aperture for the same three targets. It is important to highlight that both the AGPM and the center of the circular aperture move following the same path in both cases. To estimate the relationship between the two positions, we make the simplest assumption; that is, a constant value, which is evaluated for each observing sequence. We found no evidence of a more complex relationship between the two positions than this one (e.g., a time dependence, see Appendix\,\ref{Apx:AGPM-CA}). Those constant values are determined as;

\begin{equation}
    \Delta X = \overline{X_\mathrm{AGPM}-X_\mathrm{CA}}  \pm  \delta(X_\mathrm{AGPM}-X_\mathrm{CA})
\label{AGPM-BC_X}
\end{equation}

\begin{equation}
    \Delta Y = \overline{Y_\mathrm{AGPM}-Y_\mathrm{CA}}  \pm  \delta(Y_\mathrm{AGPM}-Y_\mathrm{CA}),
\label{AGPM-BC_Y}
\end{equation}

where $X_\mathrm{AGPM}$ and $Y_\mathrm{AGPM}$ correspond to the measured position of the AGPM in the X- and Y-axis for every individual sky frame. $\overline{X_\mathrm{AGPM}-X_\mathrm{CA}}$ and $\overline{Y_\mathrm{AGPM}-Y_\mathrm{CA}}$ correspond to the mean value of the difference between the AGPM position with respect to the center of the circular aperture, for both X- and Y-axis, while $\delta(X_\mathrm{AGPM}-X_\mathrm{CA})$ and $\delta(Y_\mathrm{AGPM}-Y_\mathrm{CA})$ are the standard deviation of those relative positions. For each science frame, knowing the position of the center of the circular aperture, and using those two equations, it becomes possible to determine the position of the AGPM with a precision better than $1$\,pixel (of the order of $0.1$\,pixels). The accuracy of 0.1\,pixels in the position of the AGPM contributes directly to the final uncertainty in the star position, and it is at the same order on accuracy as second-generation instruments (e.g. SPHERE or GPI).

\subsection{Estimating the stellar location}\label{sec:Star}

With the position of the AGPM known, it becomes easier to determine the location of the star behind it. With our approach, we can accurately determine the location of the AGPM knowing the center of the circular aperture. In Fig.\,\ref{fig:CA_vs_time} of Appendix\,\ref{Apx:CA} we furthermore show that the latter does not significantly move during an observing sequence, meaning that we are able to know the location of the AGPM for all frames, without assuming for instance that the AGPM is located at the local minimum in the sky-subtracted frames. The frames are cropped into $23 \times 23$ pixel images, centered on the AGPM position to minimize the computational time. Then, each cropped science frame is fitted with a model which has three different contributions to reproduce the torus; a positive 2D Gaussian distribution representing the imperfectly AO-corrected non-attenuated star image on the detector $V_{\mathrm{Star}}$, a negative 2D Gaussian functions accounting for the cancellation of the on-axis star light produced by the AGPM $V_{\mathrm{AGPM}}$, and an estimate of the background level $Z_0$. The latter is estimated in an annulus outside of the AGPM between $5$ and $11$\,pixels accounting for the halo generated around the AGPM, while the 2D (negative or positive) Gaussian profile take the general following form:

\begin{equation}
    V = A\frac{\mathrm{exp}\left[-\frac{1}{2}(X-\mu)^\mathrm{T} R(\theta) \times \Sigma^{-1}\times R(\theta)^{\mathrm{T}} (X-\mu)\right]}{\sqrt{(2\pi)^2 ||  \Sigma|| }},
\end{equation}\label{star-modeling}
where $A$ is the amplitude, $\mu$ is the center of the distribution (fixed to $X_\mathrm{AGPM}$ and $Y_\mathrm{AGPM}$ for the AGPM), $X$ are the pixel coordinates $[x, y]$, \mbox{\boldmath{$\Sigma$}} is the covariance matrix (a two dimensional diagonal matrix with $\sigma_\mathrm{x}^2$ and $\sigma_\mathrm{y}^2$ along the diagonal, allowing for elongated profiles), and \mbox{\boldmath{$R(\theta)$}} is a rotation matrix that depends on the angle $\theta$. For the distribution that represents the stellar contribution, the free parameters are $A_\mathrm{Star}$, $\mu_\mathrm{Star}$, \mbox{\boldmath{$\Sigma_\mathrm{Star}$}} (both $\sigma_\mathrm{x, Star}$ and $\sigma_\mathrm{y, Star}$), and $\theta_\mathrm{Star}$. For the AGPM, $\mu_\mathrm{AGPM}$ is fixed to the positions found previously using the circular aperture, and the free parameters are the amplitude $A_\mathrm{AGPM}$, \mbox{\boldmath{$\Sigma_\mathrm{AGPM}$}} (assuming that $\sigma_\mathrm{x, AGPM} = \sigma_\mathrm{y, AGPM}$ and therefore, \mbox{\boldmath{$R(\theta)$}}=$\mathbb{I}$ and $\theta_\mathrm{AGPM}$ is set to zero). One possible caveat is that for frames where the star is slightly off-centered with respect to the AGPM, our assumption that the negative Gaussian profile is centered at $X_\mathrm{AGPM}$ and $Y_\mathrm{AGPM}$ may not be entirely correct. Indeed, subtracting a negative profile in the wings of a positive profile may slightly shift the position of the minimum. Nonetheless, this remains a good first-order approximation and our approach is further tested in Section\,\ref{sec:C-test}.

\begin{figure*}
\centering
\includegraphics[width=17cm]{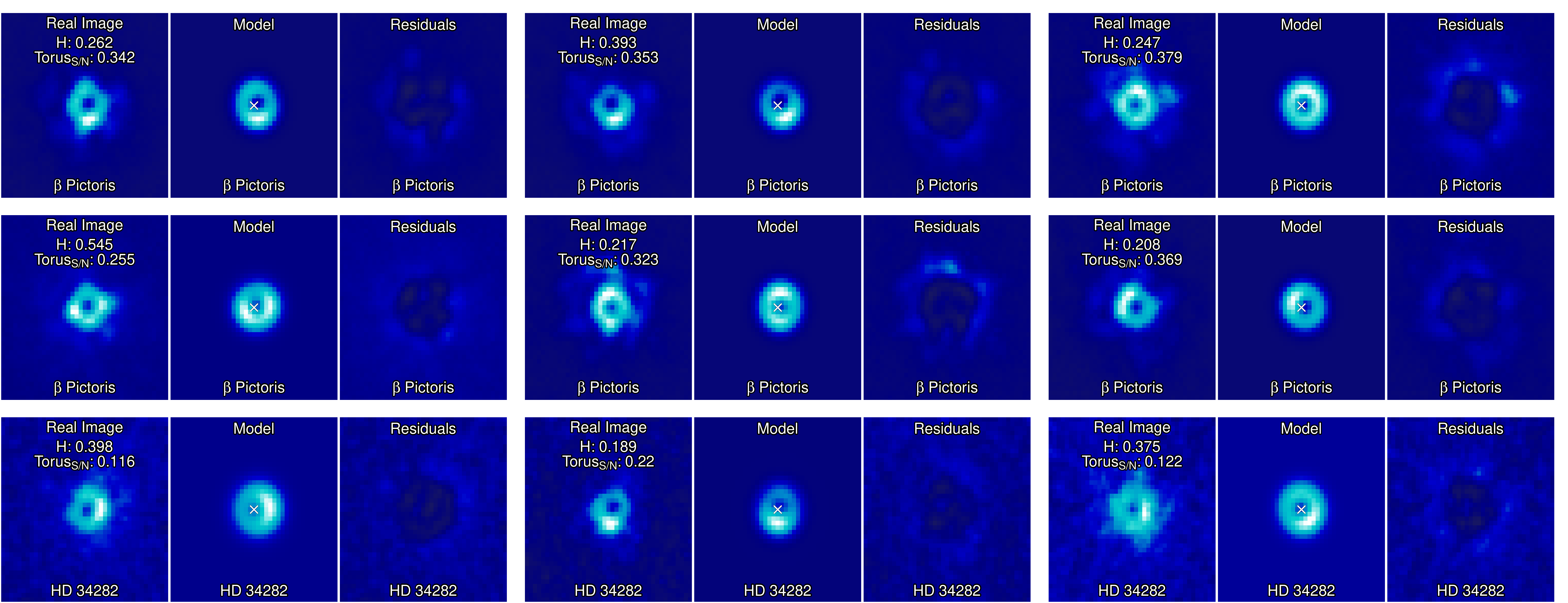}
\caption{Modeling example of sky-subtracted science images for HD\,34282 and $\beta$\,Pictoris using both the negative and positive 2D Gaussian. For each sub-panel, the observations are shown on the left, while the best-fit model is shown on the middle, and the residuals in the right. The homogeneity ($H$) and the signal to noise of the torus ($\mathrm{Torus_{S/N}}$) is labeled for each frame. The color scale is the same for each sub-panel. The white cross for each center panel marks the location of the AGPM.}
\label{fig:Real_vs_Fit}%
\end{figure*}

When minimizing the $\chi^2$ to model the sky-subtracted science frames, they are first cropped to a size of $41 \times 41$ pixels, centered on the position of the AGPM. Furthermore, we impose that the variable $\sigma_\mathrm{x/y, AGPM}$, that is, the 2D Gaussian $\sigma$ of the AGPM, cannot be greater than the $\sqrt{\sigma_\mathrm{x, Star}^2 + \sigma_\mathrm{y, Star}^2}$. We also consider the following constraints; for both the star and the AGPM $\mu$ and $\sigma$ cannot be greater than the actual size of the cropped frame. For $\sigma_\mathrm{AGPM}$, its value is constrained from fitting the sky frames, and it can vary but cannot be greater than $2.5$\,pixels\footnote{We consider the previous fit done using eq.\,\ref{eq:AGPM-center fitting} and their uncertainties to constrain the $\sigma_\mathrm{AGPM}$.}. Considering the original set of thirteen, unconstrained free parameters, we have now eight free parameters in our modelling. In the end, for each individual frame we store the location of the center of the circular aperture, as well as the eight fitted free parameters, and their associated uncertainties. Examples of the negative and positive 2D Gaussian fitting are shown in Fig.\,\ref{fig:Real_vs_Fit}, showing the sky-subtracted science frames, the models, and residuals. On the model images, we also show the location of the AGPM, showing that it does not necessarily correspond to the location of the minimum flux. Overall, with our approach, we are able to accurately find the position of the star behind the AGPM in the sky-subtracted science images with typical uncertainties of 0.20 and 0.19 pixels in the X and Y axes (see Section\,\ref{sec:C-test}).

\subsection{Frame registration and Selection}\label{FS}

The motivation behind frame registration is to characterize every individual image to identify sub-samples that worsen some aspect of the reduction to, later on, remove those frames in the datacube, performing frame selection. Depending on the objective, there are different ways of performing this registration and selection of frames that will affect the final noise distribution, the signal of the candidate and/or its position. For example, when we are interested in obtaining the position and flux of a candidate, it would be ideal to be able to maximize the signal of the object and/or minimize the residual noise of speckles. For this, one can discard those frames whose speckles and stellar PSF/halo distributions are very different from the rest of the frames, for example. Another need could be seeking to maximize the S/N of the candidate, where the idea is to minimize the noise at the same angular separation to be sure that the signal is not being considerably contaminated by speckles. To that end, it could be possible to use frames whose statistical properties of noise are very similar as a function of angular separation, discarding those whose distributions are considerably different. Or even when we only want to increase the signal of the object, eliminating frames where the signal of the object has been very diluted due to a poor AO correction or very changing weather conditions. In practice, there is no unique way to do the frame registration and selection. The most common approaches are either carried out visually when the total number of images remains small (e.g., \citealp{Samland+2017}), or using simple criteria such as the background level (see \citealp{Stolker+2019} and \citealp{Cugno+2019}), the AO performance/correction (\citealp{Hagelberg+2010}), or studying the photometric variability of the point-sources in the field-of-view \citep{Keppler+2018}.

A way to quantify the degree of similarity among frames is to characterize the torus shape described in previous sections. One way to do this is by measuring the distribution of speckles and inhomogeneities generated around the AGPM, produced by a misalignment between the star and the AGPM, the weather conditions and/or a poor AO correction, for instance. Similarly, the ring-shaped contribution (torus with isotropic distribution) formed due to the star-AGPM combination, can give us information mostly about the centering. The best way to measure both is by using image reconstructors, such as the pseudo-Zernike moments. The pseudo-Zernike moments were defined in \cite{Teh+88} as a tool to obtain information about images and their reconstructions. The pseudo-Zernike moments are defined from the (pseudo-)Zernike polynomials (\citealt{Zernike1934} ,\citealt{Bhatia+Wolf54}), and they comprise a complete set of functions orthogonal on the unit disk. \cite{Teh+88} showed that pseudo-Zerinke moments are less sensitive to image noise than the conventional Zernike moments. They are routinely used in optics, atmospheric sciences, and atmospheric turbulence corrections using adaptive optics systems (see for example, \citealp{Noll+76}, \citealp{Ma+17}, \citealp{Fusco+06} or \citealp{Sauvage+16}).

\begin{figure*}
\centering
\includegraphics[width=7.5cm]{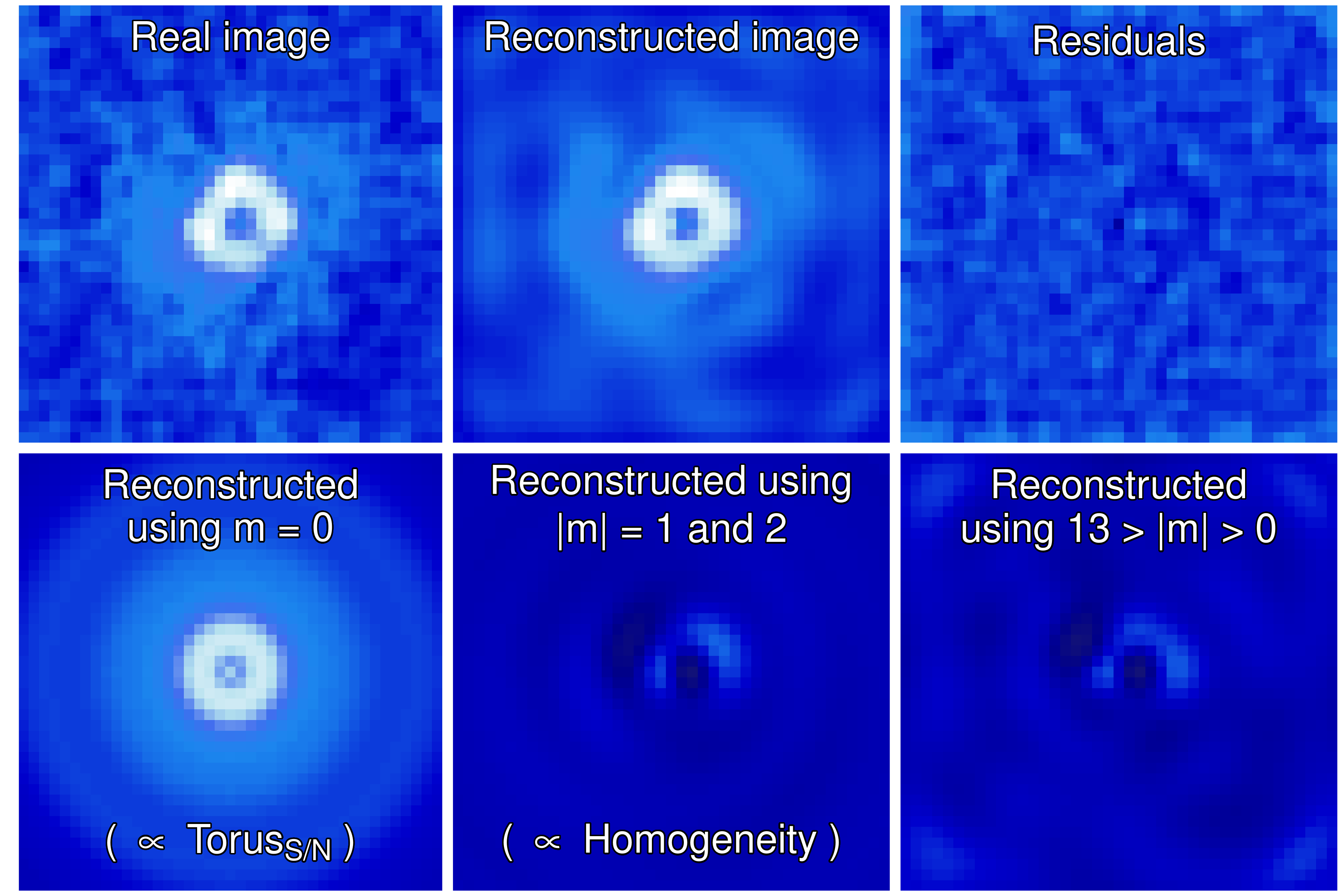}
\centering
\includegraphics[width=7.5cm]{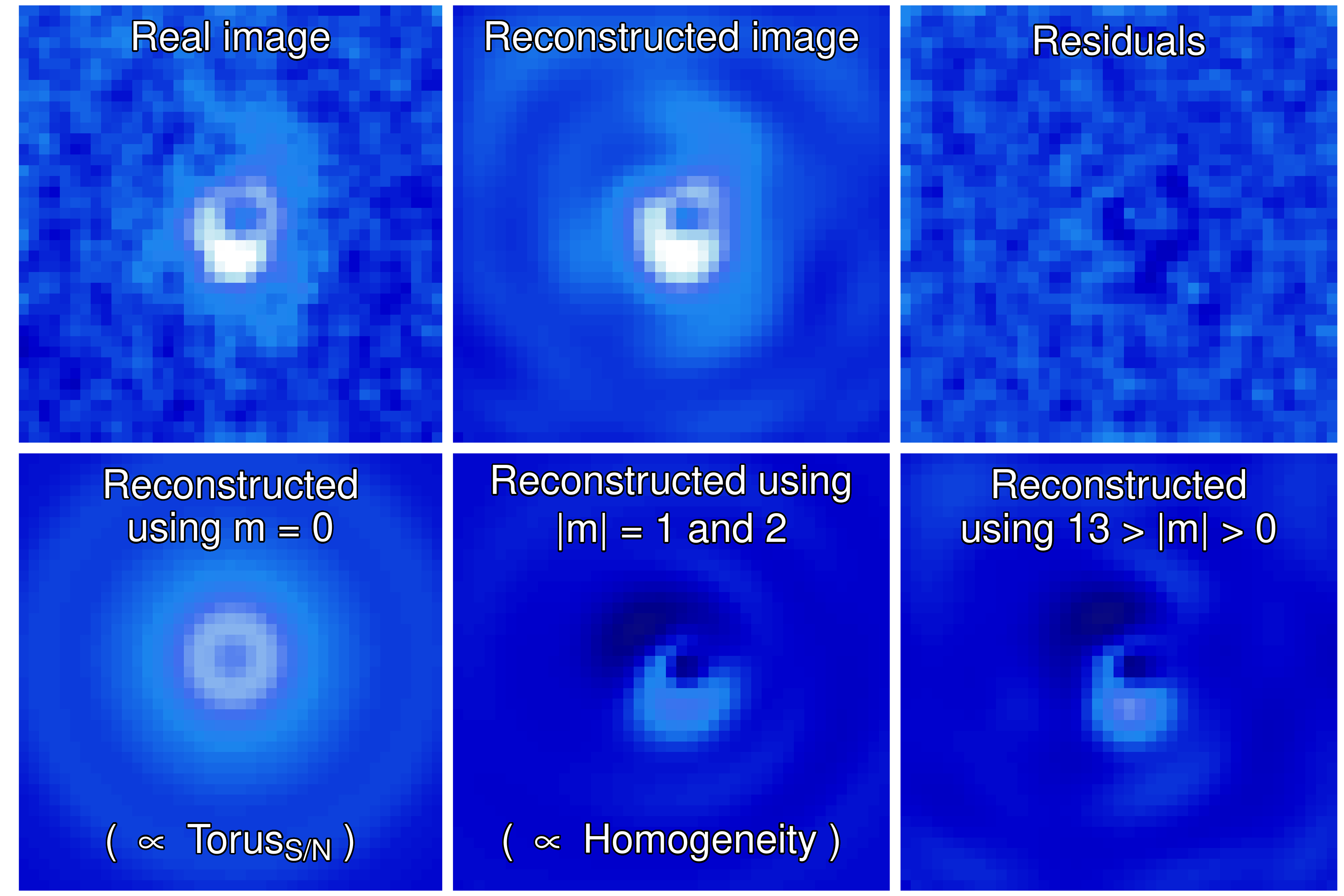}
\caption{Two examples of reconstructed images using the pseudo-Zernike moments. For each panel: \textit{Top-left:} observations; \textit{Top-middle:} reconstructed image using $m$ up to $12$; \textit{Top-right:} residuals of the modeling; \textit{Bottom-left:} reconstruction of the image using only $m=0$, related to radial and uniform azimuthal distribution; \textit{Bottom-middle:} reconstruction of the image using $\lvert m \rvert= 1$ and $2$, corresponding to the adopted inhomogeneous contribution; \textit{Bottom-right:} reconstruction of the image using  $1 \leq \lvert m \rvert \leq 12$, related to all the azimuthal contributions. For each panel, the color scale is the same (for the six images).}
\label{fig:Reconstructed_images_psZM_}%
\end{figure*}

One of the main benefits of using the pseudo-Zernike moments is its fast computational analysis when performing such decomposition, making it possible to analyze over thousands of frames quickly. We use the package \texttt{IM} \citep{IM} from \cite{R-cran} to obtain the pseudo-Zernike moments for a specific order $n$ and repetition $m$ on the sky-subtracted science frames, where $n$= 0, 1, 2, ... $\infty$ and $|m|\leq n$. The order $n$ is related to the complexity of the reference image (pseudo-Zernike moment), and $m$ to the azimuth distribution. Note that $m=0$ is purely related to radial profiles and, therefore, it is not representative of anisotropies in the reconstruction. With the coefficients obtained from the moment decomposition (hereafter, $\zeta$), it is possible to define two criteria: the homogeneity $H$ and intensity (S/N) of the torus formed around the AGPM in the science frames. The homogeneity of an image is related to the distribution of the light around the coronagraph: an image is classified with a high $H$ value when the brightness distribution is homogeneously distributed as a function of the azimuth. For each frame, $H$ is calculated combining the information of the second and third repetitions (noted as $m=\pm1$ and $m=\pm2$) of the pseudo-Zernike moments, as:

\begin{equation}
    H = 1 -\sqrt{  |\sum_{n=1}^{5} (|\zeta_{n}^{m=\pm1}|)|^2 + |\sum_{n=1}^{5} (|\zeta_{n}^{m=\pm2}|)|^2}.
\label{eq:Hom_formula}
\end{equation}
Both repetitions are related to the azimuthal brightness distribution with respect to the AGPM, therefore large values of both repetitions implies a rather inhomogeneous frame. The S/N of the intensity of the torus is calculated using the first repetition ($m=0$) of the pseudo-Zernike moments, as:
\begin{equation}
    \mathrm{Torus_{S/N}} =  \frac{\sqrt{ \sum_{n=1}^{5} (|\zeta_{n}^{m=0}|^2) }}{\sigma(25 < r < 35)},
\label{eq:SNR_formula}
\end{equation}
where $r$ is the distance from the location of the AGPM, in pixels. To estimate the S/N of the torus, its intensity is divided by the noise, which is the standard deviation measured in a ring between $25$ and $35$ pixels in radius. This region was chosen as it is free of speckles, since we want an S/N measure that reflects only the intensity of the torus and the background noise of the image not affected by residual speckles. The numerator of the previous equation corresponds to the intensity of the torus, but preliminary tests suggest that it is not by itself an ideal parameter to classify the science frames. The first repetition $m=0$ is purely related to the radial and isotropic distribution of rings and, consequently, it is azimuthally homogeneous. Therefore, any inhomogeneity of the torus would not be reflected in $\mathrm{Torus_{S/N}}$. Figure\,\ref{fig:Reconstructed_images_psZM_} shows two examples of reconstructed images and their respective residuals using the pseudo-Zernike moments. In addition to the reconstruction using only $m=0$, which corresponds to the torus intensity (without azimuthal components), the Figure also shows the reconstruction using $\lvert m \rvert= 1$ and $2$ corresponding to the homogeneity as well as the reconstruction using $1 \leq \lvert m \rvert \leq 12$. Both values, $H$ and $\mathrm{Torus_{S/N}}$, are also normalized between $0$ and $1$, for all the frames, for each object individually.

With the results from the 2D negative and positive Gaussian profile fitting as well as the information from the pseudo-Zernike moments, it is possible to further inspect the parameter space, to distinguish between the most homogeneous, well-centered frames and the bad ones. Different combinations of parameters can be used to separate the good and bad frames. However, the parameter space strongly depends on the observing conditions. Therefore, we need to adapt the selection criteria for each dataset. We find that the most crucial parameters to identify the set of good frames are the following:
\begin{enumerate}
    \item The separation between the estimated locations of both the star and the AGPM.
    \item The S/N of the torus.
    \item The background level and its standard deviation.
    \item The goodness of fit of the torus (from the 2D positive and negative Gaussian fitting).
    \item The ratio between the estimated uncertainties and their corresponding best-fit values for each of the free parameters.
\end{enumerate}
 The separation between the star and the AGPM provides a direct estimate of the quality of the centering during the observations. The S/N of the torus can indirectly provide comparable information: we find that when the S/N of the torus is large, the star is usually well-centered and the surface brightness is homogeneously distributed around the AGPM (in comparison with a misaligned case under same conditions). However, the value of the S/N is also affected by the weather conditions (e.g., clouds), decreasing the overall signal of the torus even if it is symmetric around the AGPM. The background level and the background noise level around the AGPM give us information about the strength of the speckles produced (e.g., their magnitudes and radial distribution) and, as a consequence, the quality of the AO correction. A poor AO correction usually generates many speckles, thus decreasing the contrast close to the AGPM. The ratio between the uncertainty to the estimated parameter serves as a filter to identify the frames where the fitting procedure was not successful. Other parameters, such as $H$, provide similar information and we note that their incorporation in the analysis does not necessarily improve the frame selection process (see Appendix\,\ref{Apx:H}). However, its use can be beneficial for the characterization of frames in instruments whose modeling of the torus can be very complicated or even when the torus is not clearly detected.

\begin{figure*}
\centering
\includegraphics[width=7.5cm]{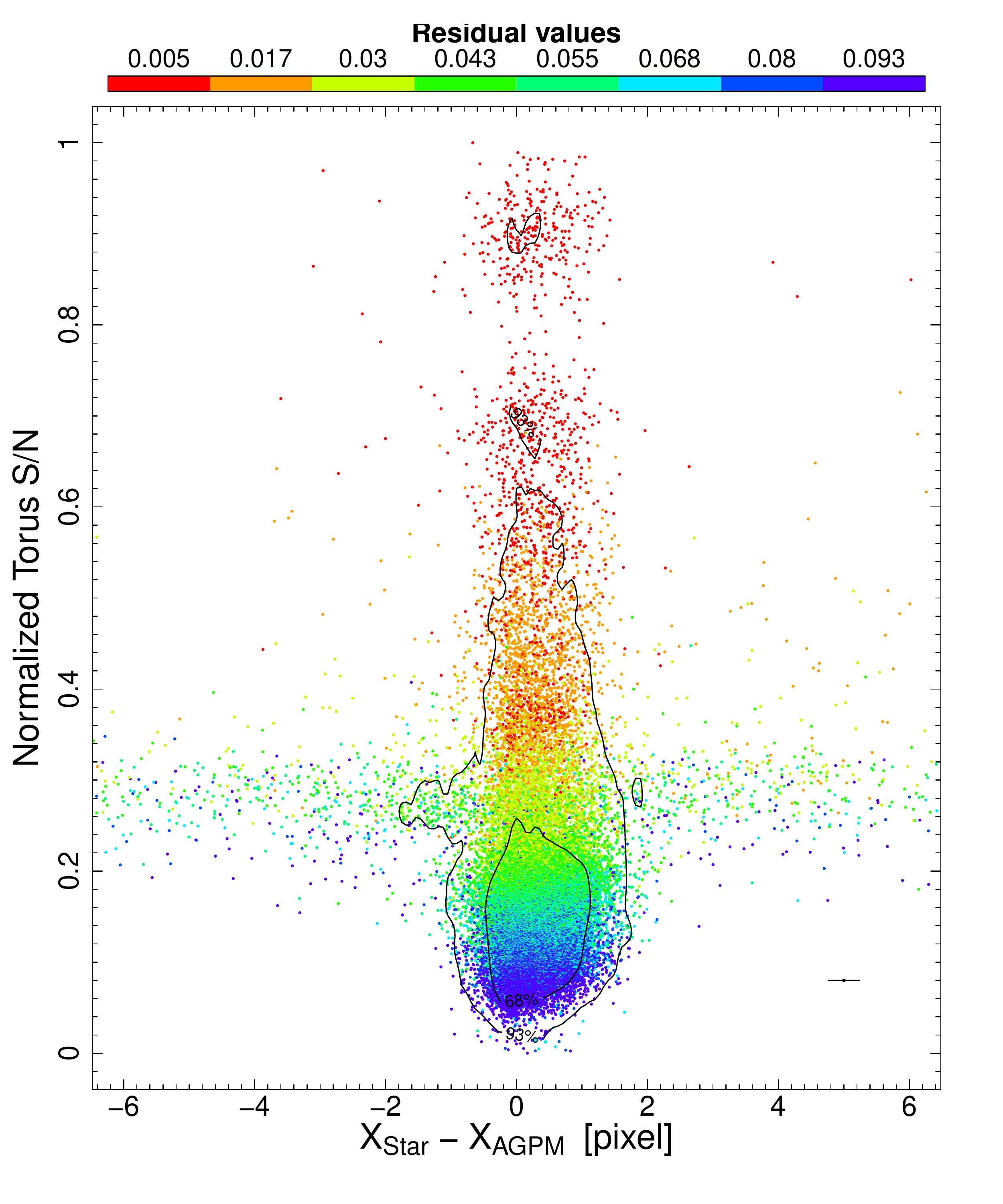}
\centering
\includegraphics[width=9cm]{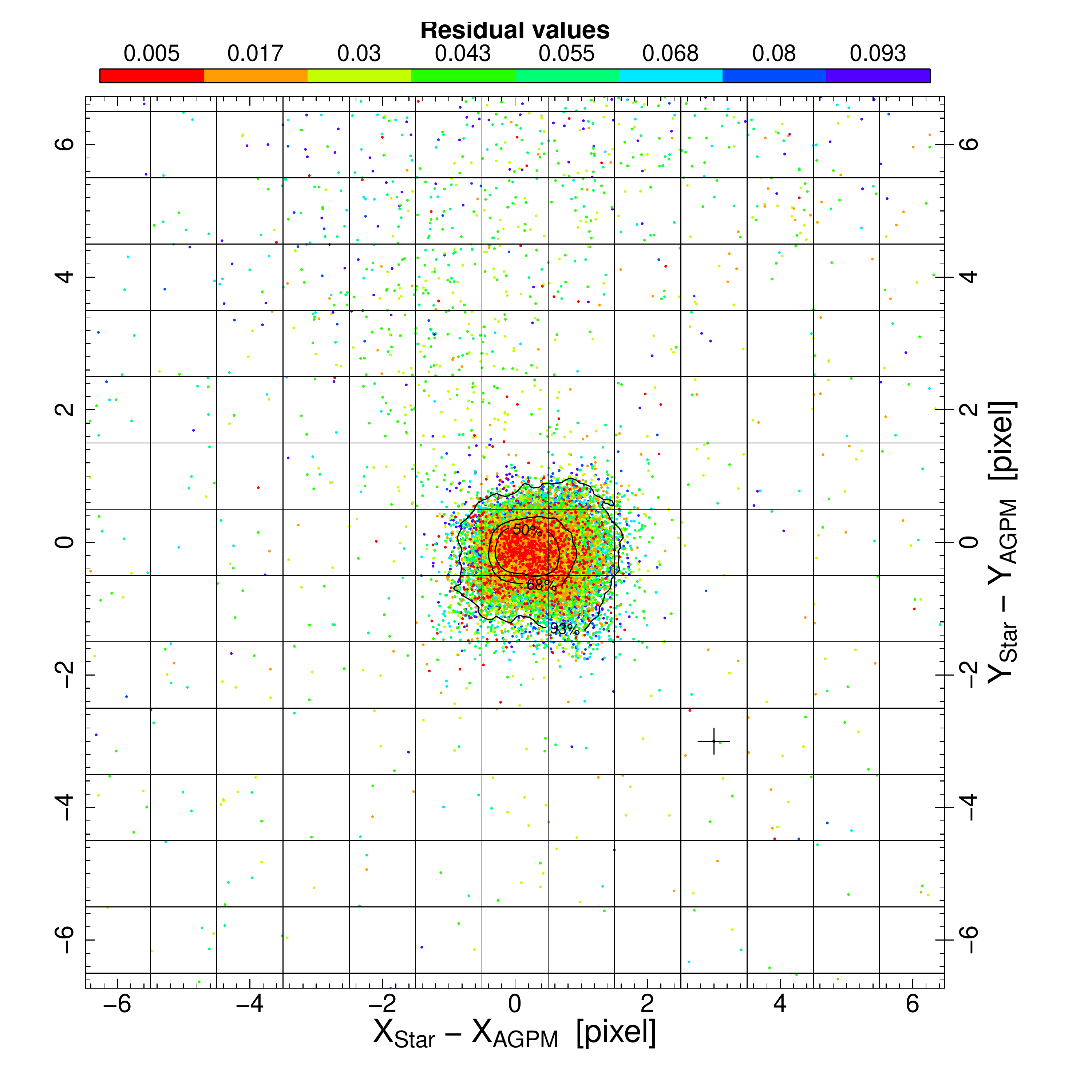}
\caption{\textit{Left}: S/N of the torus as a function of the difference between the positions of the star and of the AGPM along the X axis. \textit{Right}: relative position of the star with respect to the position of the AGPM. For both panels, the star is HD\,34282 and the color coding corresponds to the goodness of fit. The crosses marked in both panels correspond to the typical uncertainties. The contours are containing 93\% and 68\% of the data in the left panel, and 93\%, 68\% and 50\% of the data in the right panel, respectively.}
\label{fig:Example_cone_and_pos}%
\end{figure*}

Our approach for the frame selection is showcased in Fig.\,\ref{fig:Example_cone_and_pos}, for HD\,34282, showing on the left panel the S/N of the torus as a function of the separation between the star and the AGPM positions along the X-axis. The right panel shows the separation between the star and the AGPM for both axes. For both panels, the points are color-coded according to the goodness of fit when modeling the torus with two 2D Gaussian profiles (i.e., the sum of the squared residuals within the $23$ central pixels). The vertical cluster of points that appears on the left side of Figure\,\ref{fig:Example_cone_and_pos} can be explained as follows: when the star is well aligned with respect to the AGPM, the S/N of the Torus increases as the starlight is more homogeneously distributed around the AGPM. We find that in those cases, the fitting process yields better results (smaller residuals). When the separation between the star and the AGPM increases, the S/N of the torus decreases leading to larger values for the residuals. Therefore, the resulting distribution, as shown in Fig.\,\ref{fig:Example_cone_and_pos}, shows a relatively wide ring of points, with low S/N values, as well as a narrower cluster of points with larger S/N values. However, the torus is also affected by the weather conditions and the AO performance. Even if the star is well centered behind the AGPM, if the observing conditions degraded, the S/N of the torus will also decrease. The resulting distribution is therefore a mix between well or poorly centered frames, all of them affected by the observing conditions and AO performance.

\begin{figure*}
\centering
\includegraphics[width=16.5cm]{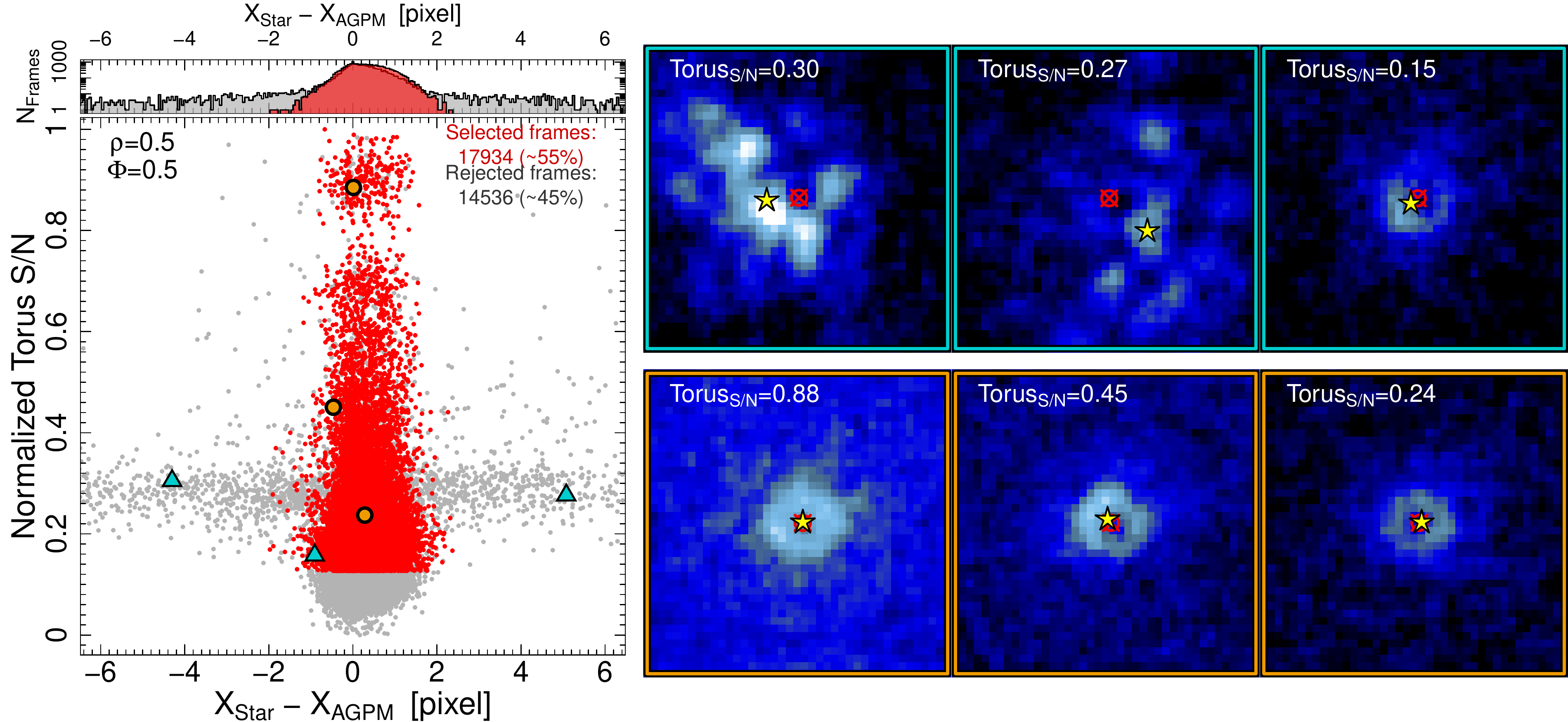}
\caption{\textit{Left}: normalized $\mathrm{Torus_{S/N}}$ as a function of the difference between the positions of the star and of the AGPM along the X axis. The color coding corresponds to the frame selected using $\phi=0.5$ and $\rho=0.5$ (red dots) and frames rejected (gray dots). The orange circles correspond to three examples of frames selected and the cyan triangles to examples of rejected frames. Top of this panel shows the histogram of both distributions in logarithmic scale. \textit{Right}: the top panel shows the three examples of rejected frames, while the bottom panel shows the examples of selected frames. The red crossed circles correspond to the position of the AGPM, while the yellow star marks the derived position of the star obtained from the negative and positive 2D Gaussian fitting. The color scale is the same for all the six images.}
\label{fig:Example_frame_selection}%
\end{figure*}

On the left panel of Fig.\,\ref{fig:Example_cone_and_pos}, the horizontal plane formed around the normalized S/N value of $0.3$ is most likely related to poor AO correction. When this happens during the observations, the background level around the AGPM increases due to the stellar leakage, and more speckles appear close to the AGPM. The algorithm can then only poorly constrain the position of the star given the overall lower quality of the AO correction. It can even find the brightest speckle next to the AGPM as the best approximation for the star location (top right panel of Fig.\,\ref{fig:Example_frame_selection}). Therefore, the intensity of the torus is dominated by the stellar leakage background, which is much larger than for other frames, as the stellar light is much more diluted and less affected by the AGPM obstruction. As a result, we observe this relatively constant level in the measured S/N when the AO correction is not working optimally. In those cases, we find that the speckles are randomly distributed in intensity and separation, and the fitting procedure finds separations up to several pixels between the location of the star and of the AGPM (instead of a fraction of a pixel).

The next step is to design the criteria that will be used for the frame selection. The goal is to select the most homogeneous frames of the sequence, the ones that share the most common features, which can then be removed using a principal component analysis (PCA) for instance. To that end, it is possible to reject the outliers in some of the distributions (e.g., star to AGPM separation, background level). This can be done by rejecting all the frames that are $5\times$MAD away from the median (MAD means the median absolute deviation). In addition, it is also possible to fine-tune the acceptance level for the S/N of the torus. This can be done by rejecting the frames that comply with being less than $\phi\times MAD\,+\,\tau$, where $\tau$ is the median and MAD is the median absolute deviation for the S/N of the torus, while $\phi$ being a free parameter which can take values from $-\infty$ to $+\infty$, in principle. It is possible to also consider the ratio between the estimated uncertainty and the best-fit value for each of the parameters used when modeling the torus with the positive and negative Gaussians. For each frame, if the ratio is larger than $\rho$ (being the second free parameter, with $\rho$ between 0 and $+\infty$) then the frame is not considered in the rest of the analysis.

Figure\,\ref{fig:Example_frame_selection} shows an example of the frame selection for HD\,34282 using $\phi=0.5$ and $\rho=0.5$, keeping $\sim55\%$ of the frames equivalent to 17\,934 frames. The left panel shows the normalized $\mathrm{Torus_{S/N}}$ as a function of the separation between the star and the AGPM location along the X axis, with the red dots corresponding to the selected frames and the gray dots to the rejected ones. The histogram at the top of the left panel shows the distribution of the selected and rejected frames. The orange circles correspond to examples of selected frames, while the cyan triangles examples of rejected ones, and those examples are shown on the right side of the Figure (bottom and top, respectively). The red crossed circles are marking the location of the AGPM, while the yellow star are located at the estimated position of the star. As was mentioned previously, the horizontal cloud of gray dots at 0.3 in $\mathrm{Torus_{S/N}}$ corresponds to frames where the AO system was not working nominally.

\begin{figure}
\centering
\includegraphics[width=9.5cm]{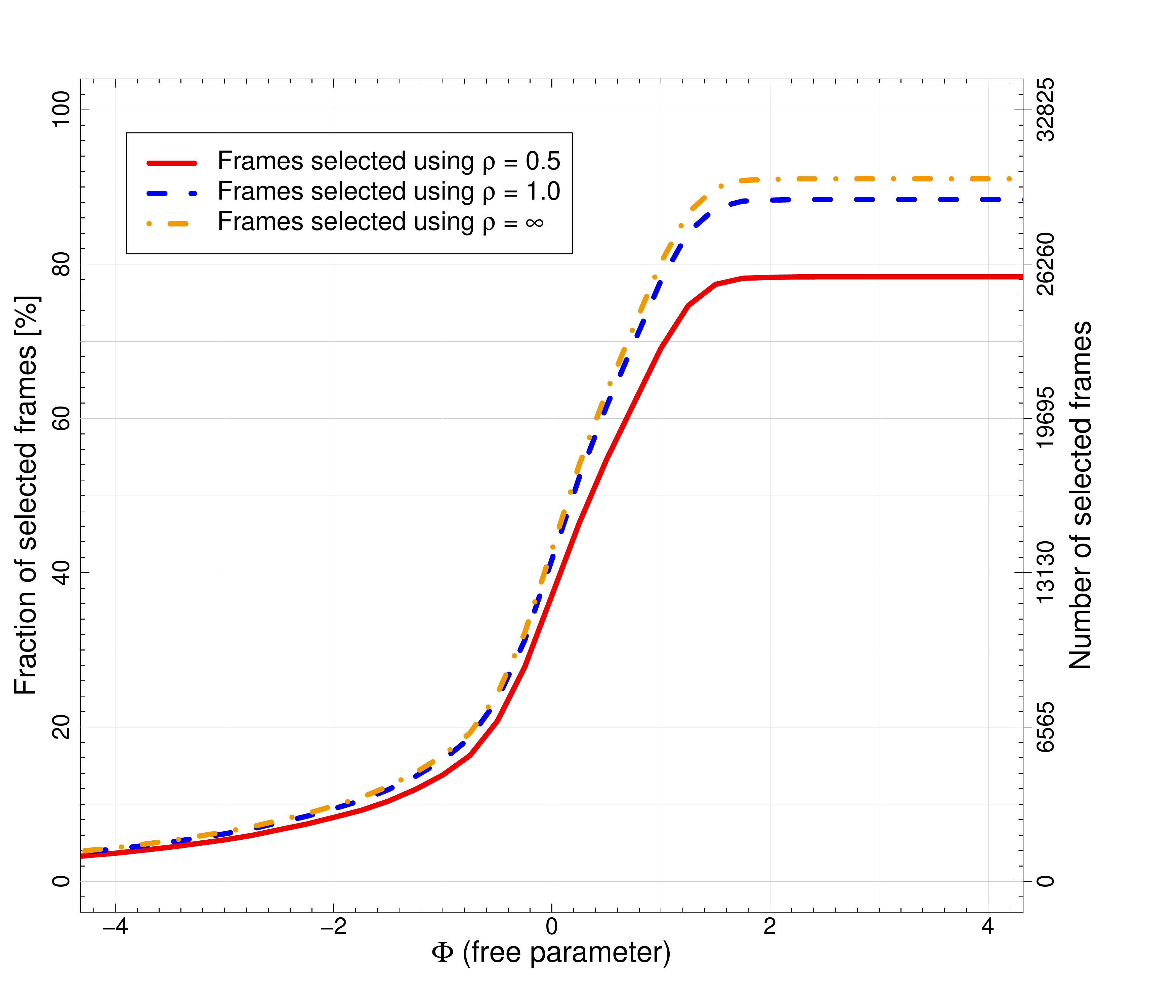}
\caption{Percentage of selected frames as a function of the free parameter $\phi$ for HD\,34282. The different lines show the behaviour for three different values of $\rho$.}
\label{fig:Perc_FS}%
\end{figure}

For our frame selection method, the most important parameter is $\phi$, as it directly relates to the percentage of frames that will later on be used in the PCA. Figure\,\ref{fig:Perc_FS} shows an example of the number of selected frames as a function of $\phi$ for HD\,34282. The parameter $\rho$ helps us to discriminate the fitted parameters that are poorly constrained and, according to Figure\,\ref{fig:Perc_FS}, it has a rather marginal effect on the number of frames selected. For this reason, we have preferred to use $\rho=0.5$ as reference value.

\section{On-sky validation: benchmarking the centering algorithm}\label{sec:C-test}

In this Section, we present a test on the robustness of our method to recover the true position of the star behind the AGPM. Given that there is no direct and model-independent way to measure the true position of the star behind the AGPM for NaCo, the best way to validate our method is to use pair(s) of stars whose angular separations are known or measurable. Therefore, we analyze observations of a target with a bright nearby known companion. The central star is HD\,104237 and its companion, HD\,104237\,B, which lies in the field of view of $\sim3.29\times3.29$\arcsec (or $121 \times 121$ pixels), is detected in all the individual science and photometric frames. Using the photometric frames, in which the central star is $>1.6\arcsec$ from the AGPM center and unsaturated, we are able to measure the separation and position angle of the companion with respect to HD\,104237. We can therefore use this information later on, to estimate the location of the central star in the science frames, correcting the position angle for changes in the parallactic angle.

The angular separation and the position angle are measured as follows. First, we fitted 2D Gaussian profiles, to obtain the positions of both point sources in the reduced photometric frames. This is done by cropping the frames to smaller windows and centering them on both components. To avoid being biased by the Airy rings, or be limited by low number statistics, we chose different sizes when cropping the images ($10\times10$, $15\times15$, and $20\times20$ pixels) and calculated the mean weighted by the respective uncertainties. Then, we measure the angular separation and position angle, propagating the estimated uncertainties on the final values. For each photometric integration, we used the median stacked image over the $400$ frames, calculating the median parallactic angle and its range as uncertainty. The photometric sequence is performed three times at the beginning and three times at the end of the observations, leading to six measurements for the angular separation and position angle. However, since the first sequence was observed with a poor AO correction (the stellar flux was completely diluted in the image), we discarded those bad quality frames. We obtained the following results: $50.69 \pm 0.20$\,pixels and $254.633\pm 0.122^{\circ}$ for the separation and position angle, respectively. With a pixel scale of $27.212 \pm 0.096$ mas\,pixel$^{-1}$ and a true North of $0.394 \pm 0.212^{\circ}$ (see \citealt{ISPY+2020}), the projected separation measured corresponds to $1.379\arcsec \pm 0.007$ and position angle of $255.027\pm 0.245^{\circ}$.

HD\,104237 (DX Cha) is an accreting Herbig Ae spectroscopic binary (A or AaAb, \citealp{Bohm+04,Garcia+2013}), with a circumstellar disk. Four stars are surrounding HD\,104237\,A within $15$\arcsec, labeled with letters from ``B'' to ``E'' (\citealt{Feigelson+2003}), forming a kind of ``mini-cluster'' (\citealt{Grady+04}). The five sources (A-E), are confirmed members of the $\epsilon$ Cha association (\citealt{Murphy+2013}), while HD\,104237\,A is the only early-type star of the association (3-5\,Myr, \citealt{Feigelson+2003}). \cite{Grady+04} reported that HD\,104237\,B (the ``source B'' in \citealt{Feigelson+2003}, or ``star-2'' in \citealt{Grady+04}), is a M\,3-4 star with an angular separation of $1.365\arcsec \pm 0.019$ and position angle of $254.586^{\circ} \pm 0.350$ with respect to HD\,104237\,A (observing date 2003-June-10). Our measurements (ISPY observations, 2017-May-16) of both position angle and angular separation (Table\,\ref{Summary_table_comp_B}) agrees well within the uncertainties with the ones presented in \citet{Grady+04}.

\begin{table}
\caption{Astrometric measurements of HD\,104237\,B.} 
\centering                      
\begin{tabular}{l c c}       
\hline\hline                
  & \cite{Grady+04} & This work   \\     
\hline\hline                        
 Obs. date & 2003-06-10 & 2017-05-16  \\
 Position angle [$^\circ$] & $254.586 \pm 0.350$ & $255.027\pm 0.245$ \\ 
 $\theta_{0}$ [\arcsec] & $1.365 \pm 0.019$ & $1.379 \pm 0.007$ \\
\hline                                   
\end{tabular}
\tablefoot{
Both the position angle and the angular separation ($\theta_{0}$) were measured with respect to HD\,104237\,A.}
\label{Summary_table_comp_B}%
\end{table}

The next step is to estimate the position of the star behind the AGPM, in a model-independent way, using the position of the companion. For each individual science frame we measure the location of the companion by fitting a 2D Gaussian profile. With the parallactic angle, the aforementioned position angle, and the angular separation, we can infer the true location of the star behind the coronagraph without any assumption about the reference center. One should note that during the observing sequence, there were some technical problems and the observations had to be stopped. As a consequence, there is a jump in the sequence, leading to two separate distributions. In addition, the observations were executed under variable weather conditions (see Table\,\ref{Summary_table}).

\begin{figure*}
\centering
\includegraphics[width=15.5cm]{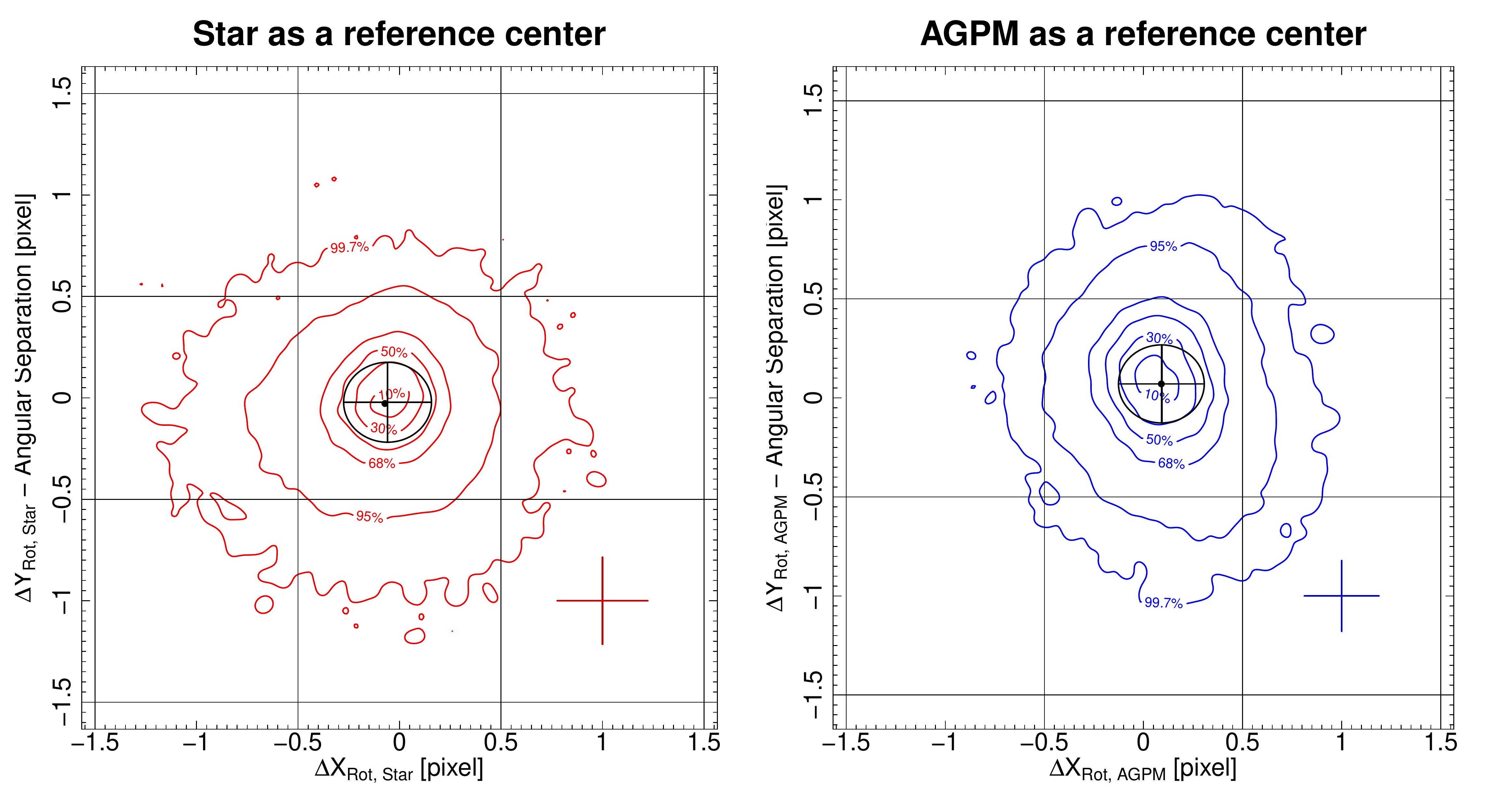}
\caption{Density contours of the position of the close companion, after de-rotating each frame by their parallactic angle and the position angle of the companion, as well as subtracting its angular separation. The de-rotation is either done at the position of the star inferred from our modeling (left) or the position of the AGPM (right). The black circle corresponds to the uncertainties on the angular separation and the position angle of the companion at the peak of the distribution. The grid shown in each panel represent pixels, and the colored crosses correspond to the typical uncertainty. The contour lines in each panel enclose the areas containing $99.7$\%, $95.5$\%, $68.5$\%, $50$\%, $30$\% and $10$\% of the sample with respect to the maximum of the distribution.}
\label{fig:Comp_Pos_Rot}%
\end{figure*}

To estimate the robustness of our method in finding the position of the star using the 2D negative and positive Gaussian profiles, we performed two different tests. The first one consists of comparing directly the position of the close companion with the previously inferred position of the star and AGPM (see Section\,\ref{sec:pipeline}). The purpose of this test is to estimate the impact of using either the star position or the center of the AGPM as the center of rotation. We de-rotate each frame to align the position of the close companion to the North using both its position angle and the parallactic angle of the frame. This is done in two separate ways: (i) using the AGPM as the rotation center and (ii) using the stellar position derived in section\,\ref{sec:Star} as the center of rotation. To center the 2D distribution around zero, we then compute the difference between the location of the companion and the location of the star (or AGPM) and subtract the angular separation of the companion measured in the photometric frames. Measuring the dispersion of the final density distribution allows us to test if the center of rotation is the correct one. For this test, we kept $\sim93$\% of the data, as we only removed the frames for which we could not determine accurately the positions of the companion and of the star (uncertainties larger than $1$\,pixel). The results are shown in Figure\,\ref{fig:Comp_Pos_Rot}. The left panel shows the density contours when the de-rotation is performed centered on the modeled position of the star, while the right panel shows the results when centered on the position of the AGPM. The median absolute deviation correspond to $0.20$\,pixels for both the X and Y axes using the star position as reference center. When using the AGPM as the reference center we measure dispersions of $0.22$ and $0.27$\,pixels. If we assume that the chosen center is correct, one would expect to obtain a distribution centered around zero with low dispersion. Since we have de-rotated the companion to the North, dispersion on the X-axis would be dominated by uncertainties on the parallactic and position angles while the dispersion along the Y-axis would be dominated by the uncertainty on the angular separation, as shown in the left panel of Figure\,\ref{fig:Comp_Pos_Rot}, where the position of the star was used as the center. However, when the position of the AGPM is used as a reference center, the distribution is elongated along the Y-axis, suggesting that the angular separation between the AGPM and the companion varies. We therefore conclude that using the AGPM as the rotation center can introduce additional biases in the reduction process, while using the  position of the science target leads to a smaller dispersion of $0.20$ pixels.

\begin{figure}
\centering
\includegraphics[width=9.3cm]{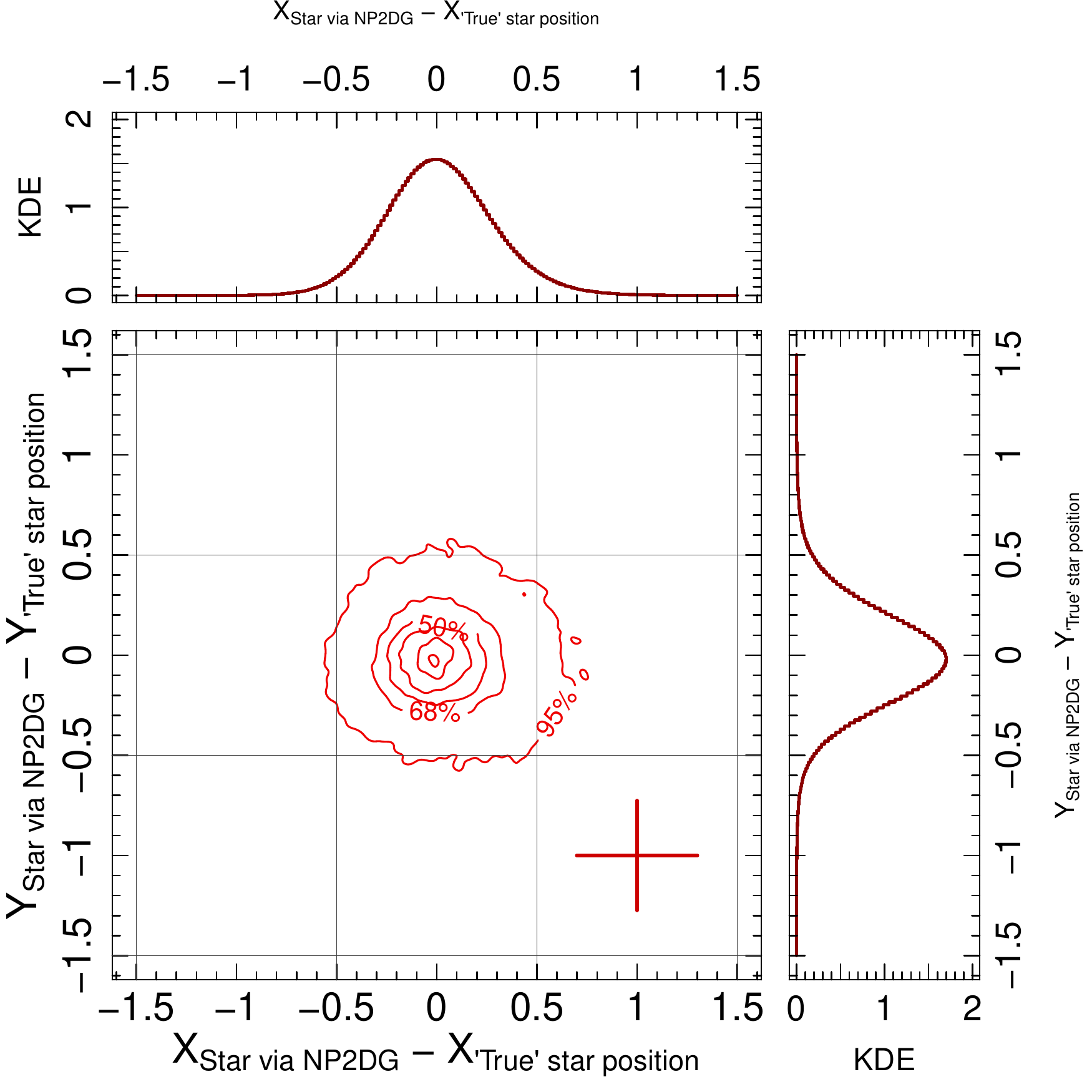}
\caption{Comparison of the star positions derived from the close companion and the ones obtained by our fitting approach (negative and positive 2D Gaussian fitting labeled as NP2DG in the labels of the figure). The top and side panels show the kernel density estimators (KDE) for the X- and Y-axes, respectively. The grid represent pixels, and the colored cross correspond to the typical uncertainty. The contour lines enclose the areas containing $95.5$\%, $68.5$\%, $50$\%, $30$\%, $10$\% and $1$\% of the sample with respect to the maximum of the distribution.}
\label{fig:True_vs_Pipeline}%
\end{figure}

The second test consists of comparing directly the true position of the central star (inferred from the position of the companion) with respect to the position of the star obtained with our method, without performing any de-rotation and therefore without having to assume a center. Figure\,\ref{fig:True_vs_Pipeline} shows the relative position of the central star using both methods. The median absolute deviation correspond to $0.21$ and $0.19$ pixels in the X and Y axes, respectively. Compared with the first test, the results strongly suggest that our method agrees well with the true position of the star within a dispersion of $0.20$\,pixel.

From both tests, we can therefore conclude that our approach to find the position of the star behind the AGPM agrees with the true position of the star estimated using its companion (within an uncertainty of $0.20$\,pixel), and at the same order of accuracy as second-generation instruments (e.g. SPHERE or GPI). \citet{Huby+15,Huby+17} demonstrated that there are some non-linear effects when the star is close to the center of the coronagraph installed for Keck/NIRC2, which we do not take into account with our approach. These effects can introduce a slight shift or increase the dispersion in our measurements on the position of the star. However, we do not observe this effect for NaCo (see Figure\,\ref{fig:True_vs_Pipeline}), or at least it is not measurable considering our uncertainties.

\section{Applications}\label{sec:Appl}

Ultimately, the goal of performing frame selection is to try to improve the contrast in the innermost regions, increasing the probability of detecting faint point sources. To quantify the effects of our centering and the frame selection approaches, we studied two targets with known companions that are located at different projected separation from their host star.

\subsection{\textbf{Signal-to-Noise ratio improvements}}\label{sec:SNR}

\subsubsection{The case of $\beta$\,Pictoris\,b}\label{sec:betaPicb}

$\beta$\,Pictoris is a young star that hosts a gas-bearing debris disk (\citealp{Smith_and_Terrile_84}, \citealp{Mouillet+97}; \citealp{Dent+14}). The star is located at a distance of $19.45\pm0.05$\,pc \citep{vanLeeuwen07}, and it is a member of the $\beta$\,Pictoris moving group with an estimated age of $22\pm6$\,Myr (\citealp{Shkolnik+2017}). $\beta$\,Pictoris hosts at least one young giant planet, $\beta$\,Pictoris\,b discovered in $2008$ via direct imaging (\citealp{Lagrange+09,Lagrange+10}). More recently the presence of another planet, $\beta$\,Pictoris\,c, has been suggested (\citealp{Lagrange+2019_c}). The planet $\beta$\,Pictoris\,b is located at $8.90_{-0.41}^{+0.23}$\,au, with a period of $20.29_{-1.35}^{+0.86}$\,yrs \citep{Lagrange+19}, and a mass of $9.3^{+2.6}_{-2.5}\,M_{\mathrm{J}}$ (see \citealp{Brandt_2021}, \citealp{Snellen_Brown_18}).

Observations with VLT/NaCo and the AGPM were carried out on February $1^{\mathrm{st}}$ 2013 (ESO program ID: $60.\mathrm{A}-9800(\mathrm{J})$). They were first presented in \cite{Absil+2013}, and re-processed in \cite{Stolker+2019}. The summary of the observations is presented in Table\,\ref{Summary_table}. It should be noted that the observations were performed under poor weather conditions, giving us the opportunity to test the effect of frame selection and centering under less than ideal conditions.

The observations were reduced as described in Section\,\ref{sec:O-DR}, including our approach for the centering of the science frames. The frame selection is performed following the workflow presented in Section\,\ref{FS}. We tried with different values for $\phi$, from $+1$ to $-1$ in steps of $0.25$. Those values, combined with $\rho = 0.5$, provide us with a wide range of frame selection efficiency, from $\sim9\%$ to $\sim65\%$ of the total number of frames ($34\,277$). With those cubes, as well as the complete dataset (i.e., without frame selection), we are able to compare our approach with other state-of-the-art post-processing techniques. In this case, we compare our results with the ones presented in \cite{Stolker+2019}. We worked with the exact same frames as in \citet{Stolker+2019} to directly compare the effect of the centering on the final images. \cite{Stolker+2019} performed a first frame selection where the frames with an unusual large background level are removed from the cube just before aligning the images. Consequently, we are using $29\,681$ frames out of the original $34\,277$ to compare the effect of different centering techniques. In addition, we also used the entire sequence of $34\,277$ frames in order to compare the effect of rejecting those frames. In the rest of the work, all images are cropped to $45\times45$\,pixels.

For each cube of data, we measure the S/N of $\beta$\,Pictoris\,b for a different number of principal components, ranging between $1$ and $120$ in steps of $1$. The signal is measured using an aperture with a radius equal to the FWHM of the PSF measured on the photometric frame, and the noise is estimated using other apertures placed at the same distance to the center of the image. To compute the S/N of the point source, we used the \texttt{python} pipeline module called \texttt{FalsePositiveModule} from the \texttt{PynPoint} package (\citealt{Amara_and_Quanz_2012,Stolker+2019}), which accounts for low number statistics (\citealp{Mawet+2014}). To better characterize the point-source in every reduced image, we fit a 2D Gaussian profile to obtain the coordinates of the companion. With those coordinates, weighted by their respective uncertainties, we estimate the mean position, to be used in the aforementioned \texttt{PynPoint} routine.

\begin{figure*}
\centering
\includegraphics[width=19cm]{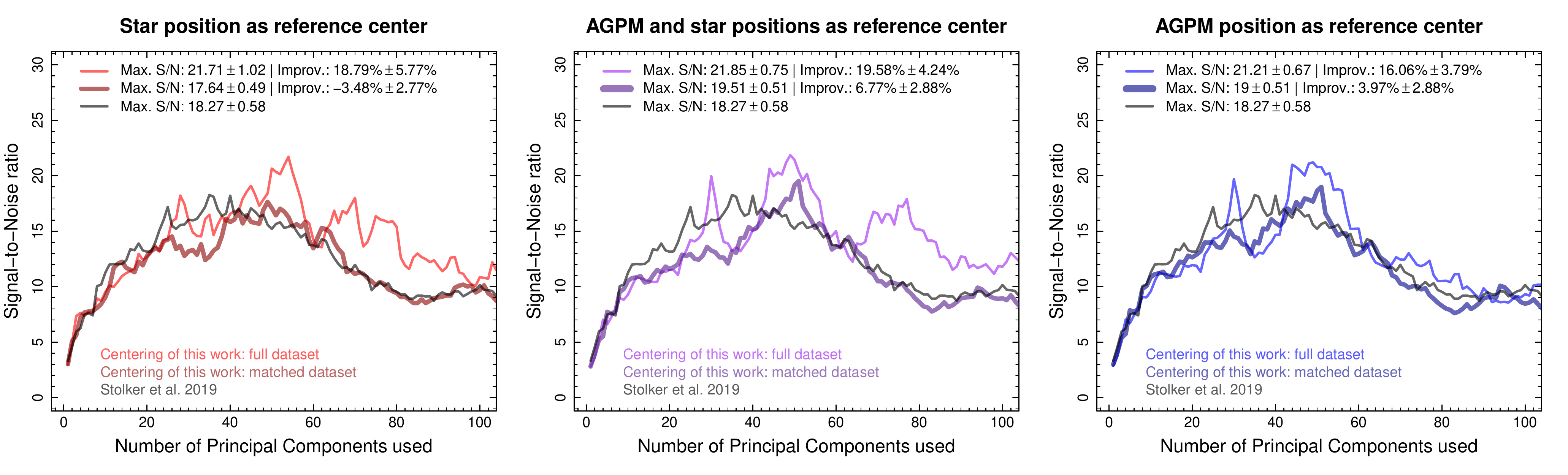}
\caption{ The S/N of $\beta$\,Pictoris\,b as a function of the number of principal components used. The solid black line corresponds to the S/N from the data reduction presented in \cite{Stolker+2019}. The thin solid lines correspond to the full dataset, while the thick solid lines correspond to the same frames used in \citet{Stolker+2019} (matched dataset), but both with our different centering options (depending on the panel). In the left panel, the frames are processed using as the reference center the star position for the full analysis. The middle panel shows the results when using the AGPM position for the principal component analysis and the star position for the de-rotation and stacking of the frames. In the right panel, the center is located at the position of the AGPM for the full analysis.}
\label{fig:BP_S/N_CT-pos}%
\end{figure*}

\begin{figure*}
\centering
\includegraphics[width=19cm]{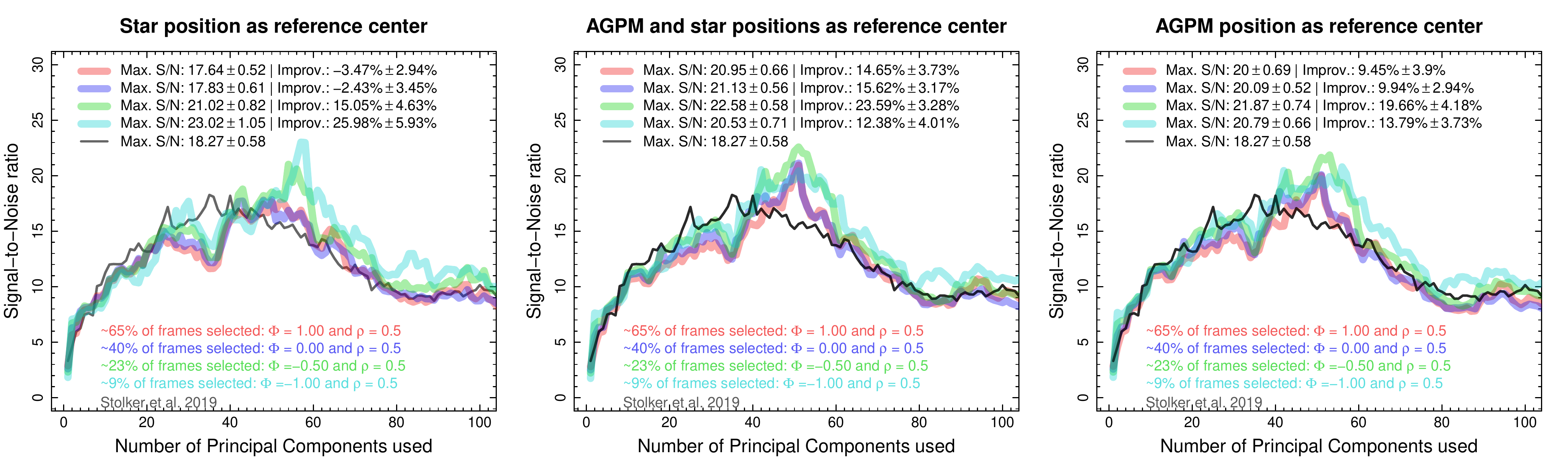}
\caption{Similar to Figure\,\ref{fig:BP_S/N_CT-pos} but this time, the thick transparent lines correspond to the reduction with frame selection using $\phi$ equal to $1$ (red), $0.0$ (blue), $-0.5$ (green), and $-1.0$ (cyan). For all the reductions, $\rho=0.5$.}
\label{fig:BP_S/N_FS-pos}%
\end{figure*}

When performing the principal component analysis, we consider three cases for the centering of the science frames; we either center them in the position of the AGPM to later be de-rotated and stacked using either the position of the AGPM or of the star as the center of rotation. We also center them at the position of the star for both the principal component analysis and the de-rotation and stacking processes. The S/N as a function of the number of principal components for the three centering strategies using the full and matched data cubes, are presented in Figure\,\ref{fig:BP_S/N_CT-pos}. The same figure, but for the frame selection with four different selection criteria is presented in Figure\,\ref{fig:BP_S/N_FS-pos}. It is worth noting that the sky subtraction for the data presented in \cite{Stolker+2019} was performed differently than in this work. They estimated the sky contribution using a principal components analysis while we are using the sky frames that are closest in time to correct each science frame. Therefore, it is not surprising that the S/N curves show some differences for certain numbers of components used, while the overall trend remains the same.

The S/N estimation does not include an estimate of its uncertainty and the measure of the S/N may vary due to the fluctuation of the noise when removing the stellar PSF and combining the images together. To obtain a first order estimate of this uncertainty, we first fit a polynomial of degree $N_\mathrm{degree}$ to the values of S/N as a function of the number of components. $N_\mathrm{degree}$ is fixed using the Bayesian information criterion (BIC, \citealp{Schwarz_78}). BIC is a criterion for model selection among a finite set of models (in this case, a family of polynomials of maximum degree $50$), where the increased complexity of the model is penalized (i.e., models with a low number of free parameters are preferred). The model with the lowest BIC value is selected, and then we subtract the resulting polygon from the S/N curve, and measure the standard deviation as an uncertainty estimator from the residuals.

Overall, we find that for $\beta$\,Pictoris\,b, we reach a better S/N (larger values) when the centering used for the principal component analysis is located at the AGPM position, followed by the de-rotation at the location of the star. This combination of centers for the entire sequence of frames provides the largest S/N, with an improvement of $19.5\%\,\pm\,4.2\%$ with respect to the analysis presented in \citet{Stolker+2019}. We note that if we compare with the matched dataset ($29\,681$ frames), with this combination of centers, we obtain a slightly larger S/N compared to the one presented in \citet{Stolker+2019}, but within the uncertainties.

Concerning the S/N when applying frame selection, we first note that the S/N agrees with the reduction presented in \citet{Stolker+2019} with an increase in the S/N between $40$ and $60$ principal components. Except for values of $\phi$ of $1$ and $0$, and using only the star position in the post-processing analysis, overall we obtain significant improvements in the S/N of $\beta$\,Pictoris\,b. The best S/N values are $23\pm1$ when using only the star as a reference center, and $22.6\pm0.6$ when using the combination of both AGPM and star position as reference centers. However, considering the uncertainties, we conclude, and in agreement with the full dataset, that the combination of both AGPM and star position provides the best reduction.

\begin{figure}
\centering
\includegraphics[width=9cm]{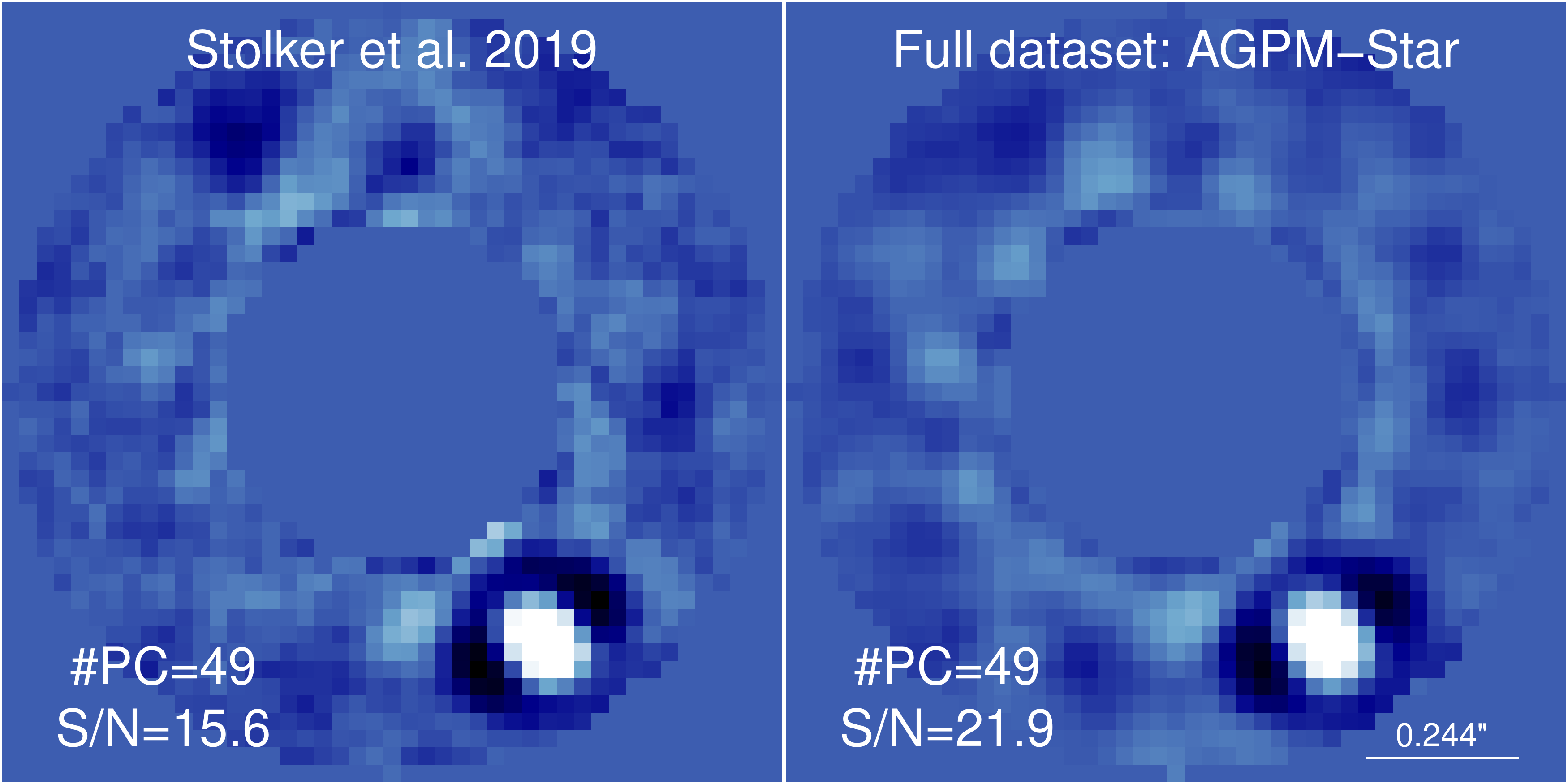}
\caption{ \textit{Left}: Images of $\beta$\,Pictoris\,b using the reduction from \citep{Stolker+2019}, using $49$ principal components. \textit{Right}: same as the left panel, with our approach for the centering using the entire dataset, using the AGPM position for PCA and the star position for the de-rotation and stacking of the frames. The color scale is linear and the same for both images.}
\label{fig:BP_images_comp}%
\end{figure}

One should note that the combination of both AGPM and star positions as the best reduction strategy is expected. When we set the center for the principal component analysis at the location of the AGPM, we are aligning the torus and the speckles together in all frames. Therefore, the speckles and stellar PSF subtraction becomes more efficient, resulting in cleaner individual images. The improvements are most notable in the innermost regions, close to the AGPM, compared to regions that are less affected by speckles and the stellar PSF. This results in larger S/N values and can also be seen in comparing the final images in Figure\,\ref{fig:BP_images_comp}. We also remark that selecting the location of the AGPM to perform the principal component analysis seems to reduce self-subtraction effect around the companion, but we do not investigate this further in this study. Our approach to perform the de-rotation centered at the location of the star is justified by the fact that the rotational center of NaCo, when performing ADI observations, is the central star according to the manual of the instrument\footnote{According with the Section 5.8, page 64 in the VLT/NACO User Manual Issue: 102}. De-rotating at a different location would yield to smearing effects on potential companions when mean- or median-stacking the dataset.

\subsubsection{The case of R\,CrA\,b}\label{Corona}

R Coronae Australis (R\,CrA - HIP\,93449) is a HAeBe star, one of the brightest in the very young, compact and obscured Coronet protostar cluster \citep{Taylor_and_Storey_84}. The star is at an early evolutionary stage and is still embedded in its dusty envelope \citep{Kraus+09}. There are some discrepancies in the estimation of the age, which varies between $0.3$ to $3.1$ Myr (e.g., \citealp{Bilbo+92}, \citealp{Forbrich+06}, \citealp{Meyer_and_Wilking_09}, \citealp{Sicilia-Aguilar+11}, among others). The distance to the Corona Australis region is estimated to $154\pm4$\,pc in \cite{Dzib+18} using mean trigonometric parallax obtained from \cite{Gaia_DR2}. \cite{Cugno+2019} and \cite{Mesa+19} identified a close companion in the envelope of R\,CrA: R\,CrA\,b. \cite{Cugno+2019} reported a mass for the companion between $0.10-0.63\,M_{\odot}$ at an angular separation of $196.8\pm4.5$\,mas (using the first epoch of NaCo $L^{\prime}$ observations taken on the 19-05-2017), while \cite{Mesa+19} estimated a mass of $0.29\pm0.08\,M_{\odot}$ with an angular separation of $184\pm4$\,mas (using SPHERE observations, K1-K2 dual band imaging, 16-08-2018). In addition, \cite{Sissa+19} reported that the star is a spectroscopic binary, with masses for the primary and secondary of $3M_{\odot}$ and $2.3M_{\odot}$, respectively, and a period of $\sim60$\,days.

The target has been observed as part of the ISPY program, and the observations were presented in \cite{Cugno+2019}. Here we used the first of the two epoch published in \cite{Cugno+2019}. It is important to note that the star saturated around the AGPM during the observing sequence (fully saturated at $\sim 19\,000$ counts, see Appendix\,\ref{Apx:frames_corona}). We reduced the observations and estimated the S/N the same way as for $\beta$\,Pictoris\,b. We used the same central mask to block the central region as in \citet[][$0\farcs05$ in radius]{Cugno+2019}, and the frames are cropped to the same size ($39\times39$\,pixels). Even though the target was observed under good weather conditions (see Table \ref{Summary_table}),  we find that for at least half of the time, the star is shifted towards a preferential direction with respect to the AGPM, suggestive of a rather poor centering (see Fig.\,\ref{fig:Rel_Pos_and_Torus_S2N_unnorm} in Appendix\,\ref{Apx:Rel_Pos_Torus}, and Appendix\,\ref{Apx:frames_corona} for examples of frames with different centering).  

\begin{figure*}
\centering
\includegraphics[width=19.0cm]{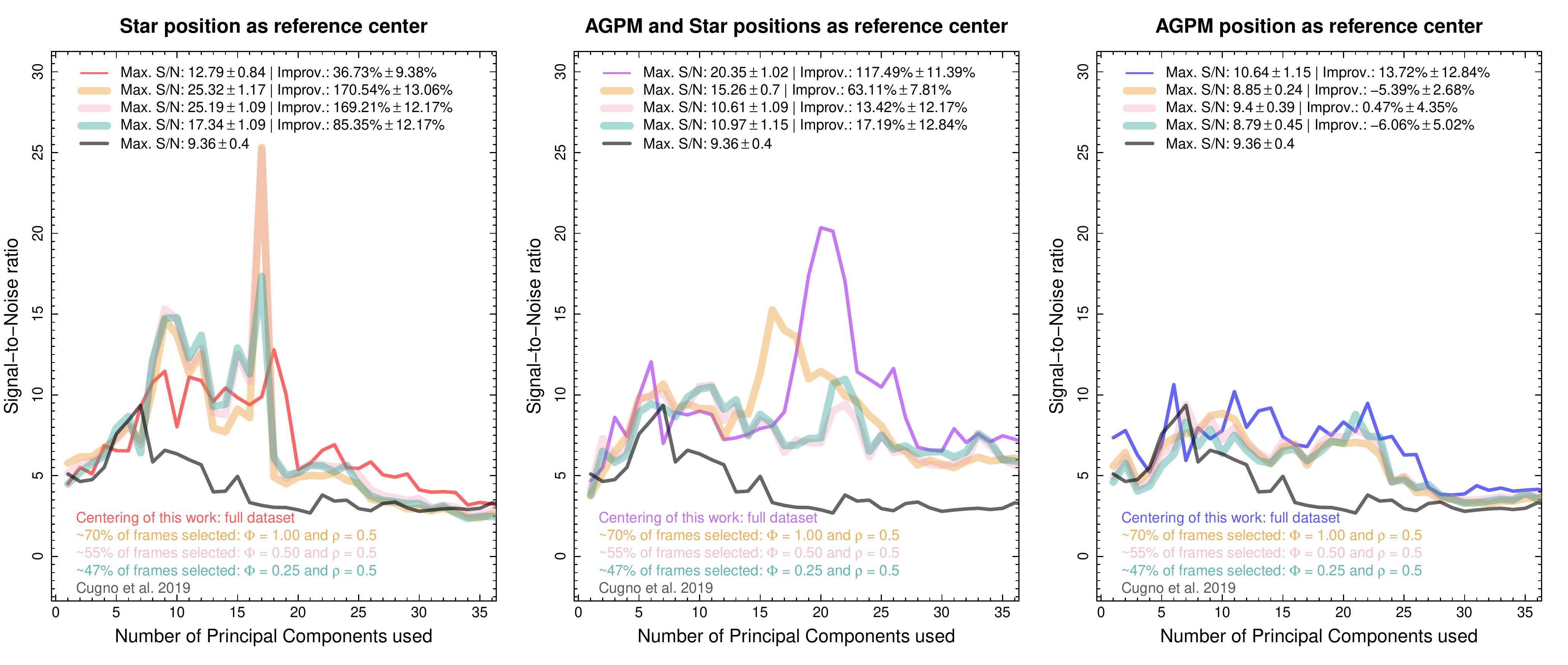}
\caption{The S/N of R\,CrA\,b as a function of the number of components used. The solid black line corresponds to the S/N from the data reduction presented in \cite{Cugno+2019}. The thin solid lines correspond to the full dataset. The thick transparent lines correspond to the reduction with frame selection using $\phi$ equal to $1$ (orange), $0.5$ (pink) and $0.25$ (cyan). For all the reductions, $\rho=0.5$. In the left panel, the frames are processed using as the reference center the star position for the full analysis. The middle panel shows the results when using the AGPM position for the principal component analysis and the star position for the de-rotation and stacking of the frames. In the right panel, the center is located at the position of the AGPM for the full analysis.}
\label{fig:RC_S/N_pos}%
\end{figure*}

We produce similar cubes as for $\beta$\,Pictoris, and we similarly consider three different centering strategies; using solely the AGPM, solely the star, or both as reference centers (see Section\,\ref{sec:betaPicb}). For each centering strategy, we produce data cubes for the full dataset, or for subsets after performing frame selection. The frame selection was performed using $\phi$ values between $1$ and $-1$ in steps of $0.25$. In the end, a total of $30$ different cubes are built, on top of the one presented in \citet{Cugno+2019} where the authors used a different centering strategy. We produce the final images using a number of principal components between $1$ and $40$ in steps of $1$, for each datacube, since for large numbers of components self-subtraction can become important. Then, the S/N of the companion is estimated for all the different reductions the same way as described for $\beta$\,Pictoirs\,b in Section\,\ref{sec:betaPicb}. In the end, we obtain $31$ different S/N curves, one per datacube, and Figure\,\ref{fig:RC_S/N_pos} shows a selection of them for the full dataset and when performing frame selection (keeping $\sim70\%$, $\sim55\%$ and $\sim47\%$ of all frames).  

From Figure\,\ref{fig:RC_S/N_pos} we highlight that using the AGPM as the center for the entire process (right panel), yields the worst S/N curve, with a maximum S/N of the order of $10$ whether we use frame selection or the full dataset (similar values were obtained by \citealt{Cugno+2019}). In the other two cases (either using the position of the star only, or both the positions of the AGPM and of the star), we note significant improvements on the estimated S/N. For all the cases, we find that the S/N is larger in the principal component range of $13$ to $25$. In particular, when we use both the AGPM and the star positions in the post-processing analysis, the peak broadens, reaching a maximum S/N of $20.3$ when using the full dataset (middle panel of Fig.\,\ref{fig:RC_S/N_pos}). This broadening of the peak suggests that not only is the noise lower, but also that self-subtraction effects become less significant as the number of principal components increases. In contrast, when using only the star as the center for the entire process, we find that the S/N of the companion is overall lower, suggesting that the PSF subtraction is less adequate in this case, despite a narrow peak when using $17$ principal components. Given the narrowness of this peak, the result should be taken with caution.

The negative patterns, due to self-subtraction effects, appear more symmetric when we use only the star position or both the star and AGPM positions as the center (see Figure\,\ref{fig:RC_images_comp} to compare the images with the best S/N). In Appendix\,\ref{Apx:Rel_Pos_Torus}, we show the distribution of relative positions between the star and the AGPM (for all the targets), and as mentioned before, the distribution for R\,CrA is not centro-symmetric with respect to the position of the AGPM. If only the AGPM is used as the center of reference, for instance, the signal of the object might be more diluted in the final image, due to the choice of center for the de-rotation. For R\,CrA, we conclude that this is the most critical point when post-processing the observations, as it can significantly improve the results in the inner regions. Regarding the impact of frame selection, we find that overall, the improvements are, at best, marginal for this specific target.

\begin{figure}
\centering
\includegraphics[width=9cm]{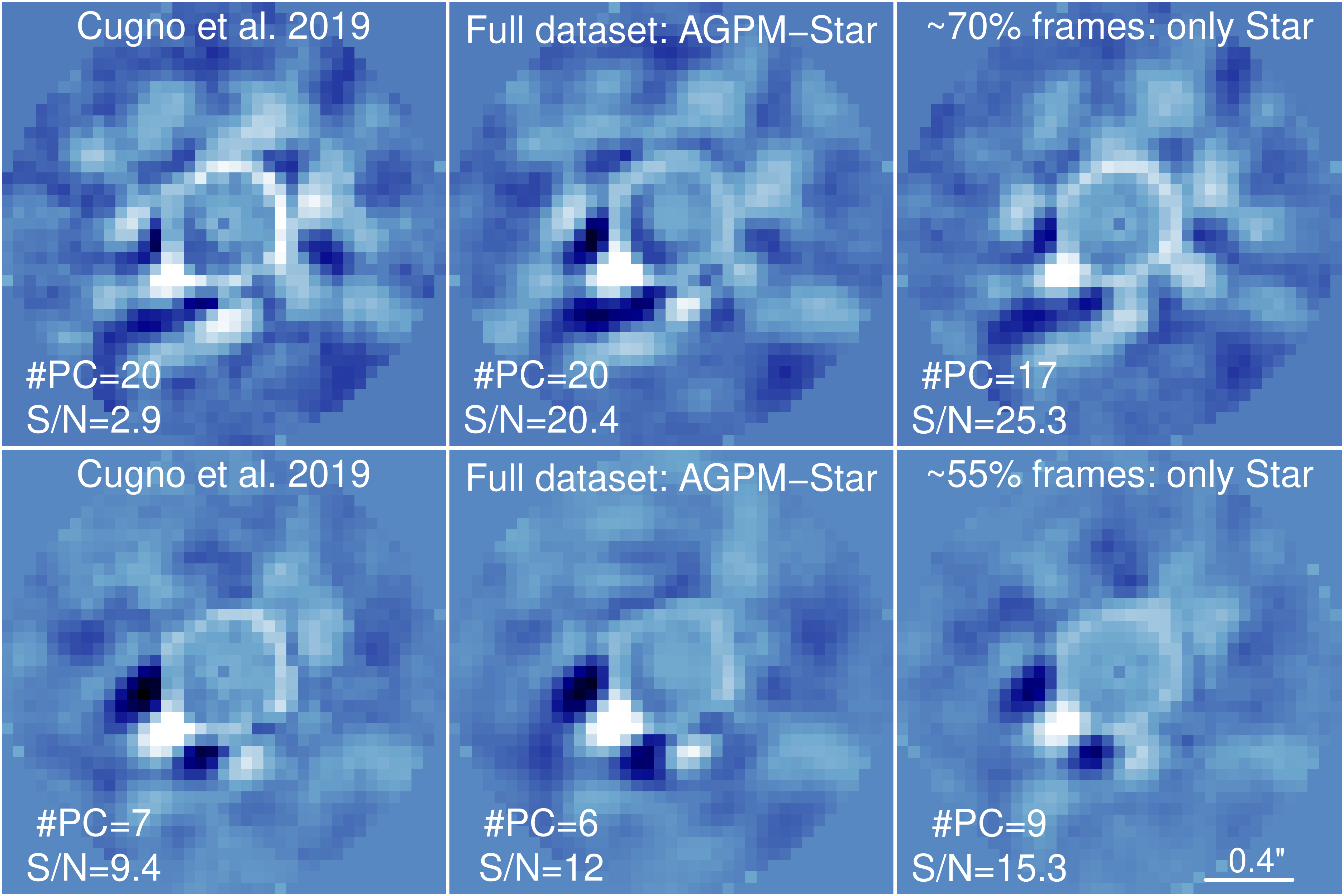}
\caption{ \textit{Top panel}: Images of R\,CrA\,b for the best S/N upper the 13 components used. \textit{Bottom panel}: Images of R\,CrA\,b for the best S/N for those values lower 13 components applied. In left the best S/N for \cite{Cugno+2019}, in middle for the entire dataset using our centering approach with the combination of the AGPM and star positions, and right the frame selection using only the star position as reference center. The top panel and the bottom panel have different color scales each one, but consistent in each panel.}
\label{fig:RC_images_comp}%
\end{figure}

\subsection{\textbf{Effects on astrometric measurements}}\label{sec:ASTR}

Astrometric measurements are critical for assessing the co-moving nature of candidate companions. The astrometric position as a function of time allows us to study the orbital motion of giant gaseous planets around their host stars, helping to better constrain their orbital parameters and masses (see, for instance, \citealt{Wang_2018,Nielsen_2020,Brandt_2021}). For this reason, we here attempt to quantify (at least, at first order), the effect of the centering and the frame selection on the astrometry of point sources.

The positions of $\beta$\,Pictoris\,b and R\,CrA\,b can be obtained following different approaches. For example, the negative fake companion technique (\citealt{Lagrange+10,Marois+2010}), which consists of injecting a negative signal at the approximate location of the source of interest in the frame sequence at each parallactic angle. The technique aims at canceling out the companion as well as possible in the final post-processed image and has been used for instance in \cite{Stolker+2019} and \cite{Cugno+2019} (see Sec. 3.5.2 of \citealt{Stolker+2019} for suitable information). There are more methods to determine the astrometric position of companions, for example using the ANgular Differential OptiMal Exoplanet Detection Algorithm (ANDROMEDA, \citealt{Cantalloube_Mouillet_Mugnier+2015}), or the Vortex Imaging Processing (VIP, \citealt{Gomez-Gonzalez+2016}), which have been applied in \cite{Biller+2021,Jorquera+2021,Langlois+2021}. However, all these methods require considerable computing power, proportional to the size of the field-of-view and the number of frames. In the case of NaCo, we have $10\,000-30\,000$ frames per target, making these procedures very challenging to carry out. \cite{Stolker+2019} and \cite{Cugno+2019} addressed this problem by stacking several images together, reducing the number of frames from thousands to hundreds. Given the fact that when using frame selection we remove frames that are not necessarily consecutive in time, we cannot easily stack the images in a similar manner, and we therefore opted for a simpler approach; fitting 2D Gaussian profiles to the images.

For each dataset ($\beta$\,Pictoris and R\,CrA, with or without frame selection, different centering, or original reductions), we computed the final images using a different number of principal components (between 1 and 35 for R\,CrA, and between 1 and 120 for $\beta$\,Pictoris). For each image, we then fitted a 2D Gaussian profile to find the location of the companion. The free parameters are the amplitude, center, rotation angle and width for both axis. For each dataset, we then computed the mean value for the X and Y positions, weighted by their corresponding uncertainties. Residual speckles systematically affect the position of the companion in an image using a certain number of components. However, this residuals change while increasing the number of components used, affecting the observed position in a different manner.
In this way, we can obtain a better approximation of the position of the companion (least affected by this systematic), by calculating the median between all the positions obtained using a wide range of principal components.

\begin{figure*}
\centering
\includegraphics[width=14cm]{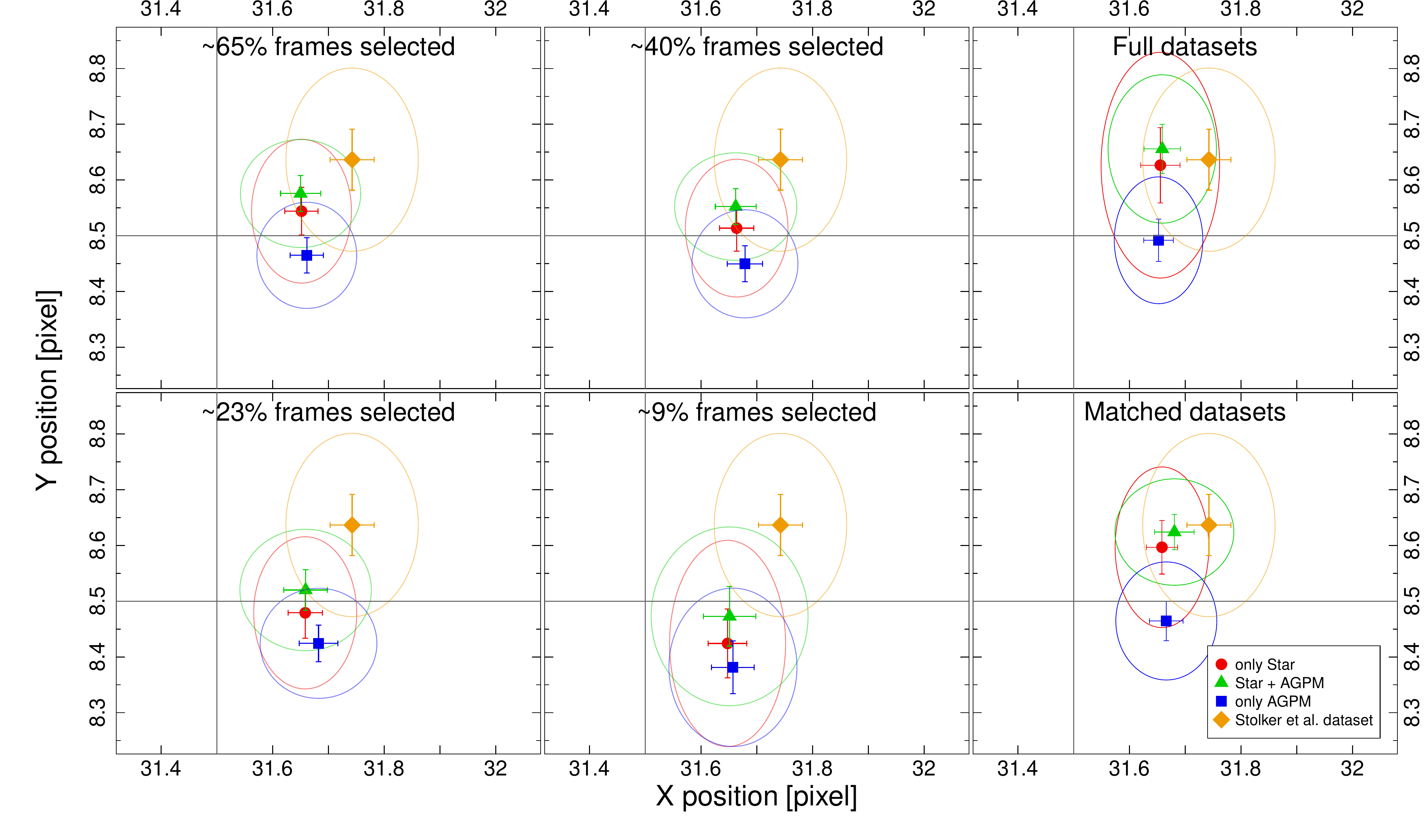}
\caption{ Position in the final image (45$\times$45 pixels) of $\beta$\,Pictoris\,b. Each sub-panel corresponds to a specific dataset (frame selections, full dataset and matched dataset). The solid red circles correspond to the datasets that use the position of the star as the reference center, while the solid blue squares use only the AGPM, and the solid green triangles both (star + AGPM). The orange diamond corresponds to the reduction of \cite{Stolker+2019}. The ellipses correspond to 3-$\sigma$ uncertainties, while the error bars 1-$\sigma$ uncertainties. }
\label{fig:BP_pos_comp}%
\end{figure*}

\begin{figure}
\centering
\includegraphics[width=9cm]{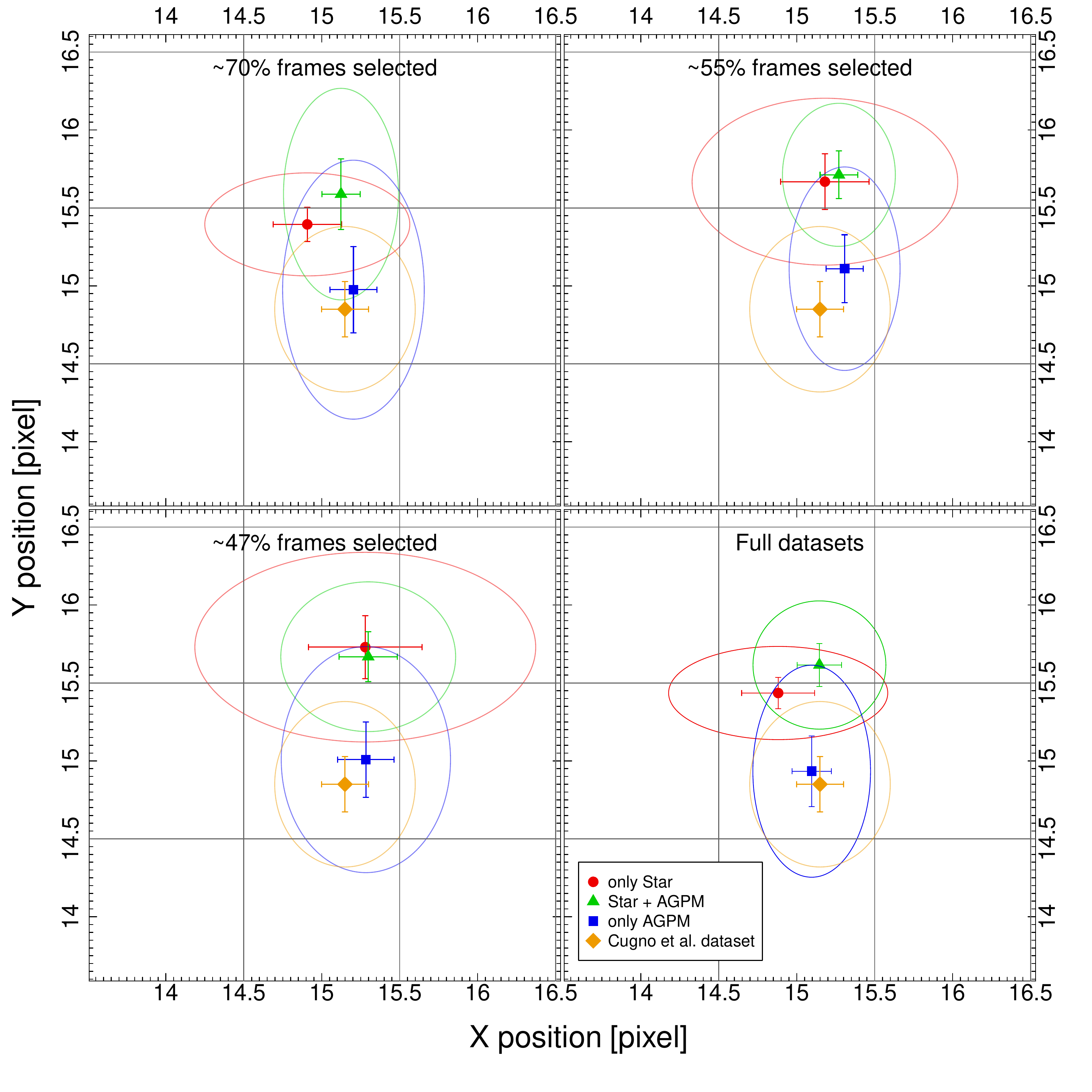}
\caption{ Similar to Fig.\,\ref{fig:BP_pos_comp} but for R\,CrA\,b (compared with \citealt{Cugno+2019}).}
\label{fig:RC_pos_comp}%
\end{figure}

Figures \ref{fig:BP_pos_comp} and \ref{fig:RC_pos_comp} show the results for $\beta$\,Pictoris\,b and R\,CrA\,b, where we plot the results for different datasets. Despite small differences, the astrometry remains consistent for all the datasets on each companion (including the dataset from \citealt{Stolker+2019} and \citealt{Cugno+2019}) at the $3\sigma$ level. The Figures show that, overall, frame selection reduces in some cases the uncertainties on the astrometric measurements, when compared to the reductions for the full datasets. For the astrometry of R\,CrA\,b, the dataset from \cite{Cugno+2019} agrees better with our solution using only the AGPM. The same is observed when comparing the S/N where both follow the same trend (see Fig.\,\ref{fig:RC_S/N_pos}). The astrometry uncertainties for R\,CrA\,b are larger than for $\beta$\,Pictoris\,b and, even though all the datasets are consistent with each other, differences of $\sim0.7$ pixels are registered corresponding to the magnitude of the asymmetry on the distribution of the differences between star and AGPM positions. Overall, we do not observe significant difference of the final astrometry at $3\sigma$ level. This is the case, even without accounting for additional uncertainties such as on the pixel scale (typically $27.2 \pm 0.1$\,mas/pixel) or the True North ($0.4 \pm 0.2$ deg.). We therefore conclude that the centering approach seems to primarily have an impact for the S/N of the detection, and can reduce the uncertainties associated with the astrometry but marginally affects the astrometric measurements themselves.

\section{Summary and conclusions}\label{sec:concl}

In this paper, we presented the new CenteR algorithm to improve the performance of coronagraphic NaCo observations at $L^{\prime}$ band. One of the fundamental challenges of such observations is to find the location of the star behind the coronagraph (AGPM). Our method makes use of the sky frames to determine the location of the AGPM in the science frames, thus alleviating the degeneracy when modeling the contributions from the central star and AGPM to infer the positions of either. The method described in this paper can be applied to any instruments for which the circular aperture of the coronagraphic mask is visible in all the frames and the AGPM emits thermally. We also presented an in-depth analysis of the impact of frame selection when dealing with datasets containing tens of thousands of frames. Our approach is based on the modeling results of the star and AGPM as well as the pseudo-Zernike moment decomposition of the sky-subtracted frames to select the most homogeneous images. We also investigated the impact of choosing different centers for the principal component analysis and the de-rotation of an angular differential imaging sequence. Our pipeline is tested on two different stars harboring known companions at short and intermediate separations (R\,CrA and $\beta$\,Pictoris). We also tested our pipeline on a star that hosts a bright circumstellar disk (HD\,34282, see Appendix\,\ref{Apx:Disk}). However, the analysis of the disk and its detectability and signal-to-noise improvements is beyond the objectives of this article.

Using observations of a star with a bright nearby companion, HD\,104237, we demonstrated that our approach can successfully recover the true position of the star behind the AGPM, with a $\sim 0.20$\,pixels dispersion, even if the observing sequence is interrupted. Since our method provides both the positions of the AGPM and of the star, we showed that performing the PCA centered at the location of the AGPM followed by the de-rotation at the location of the star yields better results compared to using only one single position (either the star or the AGPM). Performing PCA at the location of the AGPM removes the diffracted starlight and speckles more efficiently. But the de-rotation and stacking of the frames should be performed centering at the location of the star, to avoid smearing the flux from any nearby point sources. For NaCo, the center of rotation coincides with the science target, so that no shifts have to be applied beforehand. We find that for point sources, the flux is less smeared, yielding better detections and larger S/N values. Regarding the astrometry, we find that all the centering strategies are consistent with each other within $3\sigma$, while the corresponding uncertainties are smaller when using both the star and the AGPM positions. We compared our reductions with published results, reduced with state-of-the-art pipelines, and found that we are able to improve the S/N of the detection of point sources around $\beta$\,Pictoris and R\,CrA by $24\pm3$\,\% and $117\pm11$\,\%. This may also be related to the intrinsic quality of the centering during the observations. Our method yields larger S/N values when the star is poorly centered behind the AGPM and the distribution of the difference between the star and AGPM positions is not centro-symmetric. In contrast, our method reaches similar or slightly better S/N values when the distribution of the difference between the star and the AGPM positions is centro-symmetric.

To the best of our knowledge, this study is the most thorough investigation of the impact of frame selection for datasets consisting of thousands of images. We concluded that frame selection, at best, marginally improves the final reduction. We find that frame selection is most useful when either the observing conditions or the AO correction degraded during the observing sequence. The frames taken under such conditions would dilute the astrophysical signal and removing them does help decreasing the noise, especially in the innermost regions, more prone to be dominated by speckles. Regarding the astrometry, we found that frame selection can slightly decrease the uncertainties depending on the centering strategy and the percentage of frames kept, but overall provide consistent solutions when comparing with the full datasets. On the other hand, in cases for which the centering is poor and anisotropic (i.e., the star moved in a preferential direction during the observing sequence), the centering strategy using the full dataset can have an impact on the astrometry. R\,CrA is a clear example of this situation, where we estimated a possible bias in the position of about 0.7 pixels or $\sim$19\,mas when using only the AGPM centering in the entire processing. Nonetheless, for this particular case, the different centering strategies implemented provide results that are consistent with each other within $3\,\sigma$. 

There is a wealth of archival data obtained with an AGPM vector vortex coronagraph. First and second generation instruments that use an AGPM such as VLT/NaCo (\citealt{Rousset+2003};\citealt{Lenzen+2003}), LBT/LMIRCam (\citealt{Skrutskie_2010}; \citealt{Defrere_2014}) or Keck/NIRC2 (\citealt{Vargas_Catalan_2016}; \citealt{Serabyn_2017}) and its previous and ongoing surveys (for example, LEECH \citealt{Skemer_2014}; \citealt{Stone_2018}), may benefit from a re-analysis using the method described in this study. New generation instruments, VLT/ERIS or ELT/METIS among others, may benefit from the use of an AGPM and the centering method and techniques described in this study. Thanks to the improved centering of the whole dataset, our approach yields better results closer to the AGPM, in the typical range $10-50$\,au from the star, where the population of giant planets is still poorly constrained.

\begin{acknowledgements}

The authors thank the anonymous referee for a thorough review. The comments helped improving the manuscript, especially with respect to the astrometric measurements, but also the overall structure of the paper. N.~G., J.~O. and A.~B. acknowledge financial support by ANID, -- Millennium
Science Initiative Program -- NCN19\_171. N.~G. acknowledges grant support from project \emph{CONICYT-PFCHA}/Doctorado Nacional/2017 folio 21170650. J.~O. acknowledges financial support from the Universidad de Valpara\'iso, and from \emph{FONDECYT Regular} (grant 1180395). A.~B. acknowledges support from \emph{FONDECYT Regular} 1190748. A.~M. and A.~Q. acknowledge the support of the DFG priority program SPP 1992 ``Exploring the Diversity of Extrasolar Planets'' (MU 4172/1-1). Th.~H. acknowledges support from the European Research Council under the Horizon 2020 Framework Program via the ERC Advanced Grant Origins 83 24 28. G.~M.~K is supported by the Royal Society as a Royal Society University Research Fellow. T.~S. acknowledges the support from the ETH Zurich Postdoctoral Fellowship Program. S.~P.~Q. and G.~C. thank the Swiss National Science Foundation for financial support undergrant number 200021\_169131.

\end{acknowledgements}

\bibliographystyle{aa}
\bibliography{references}

\begin{appendix}

\section{Motion of the circular aperture as a function of time}\label{Apx:CA}

\begin{figure*}
\centering
\includegraphics[width=16cm]{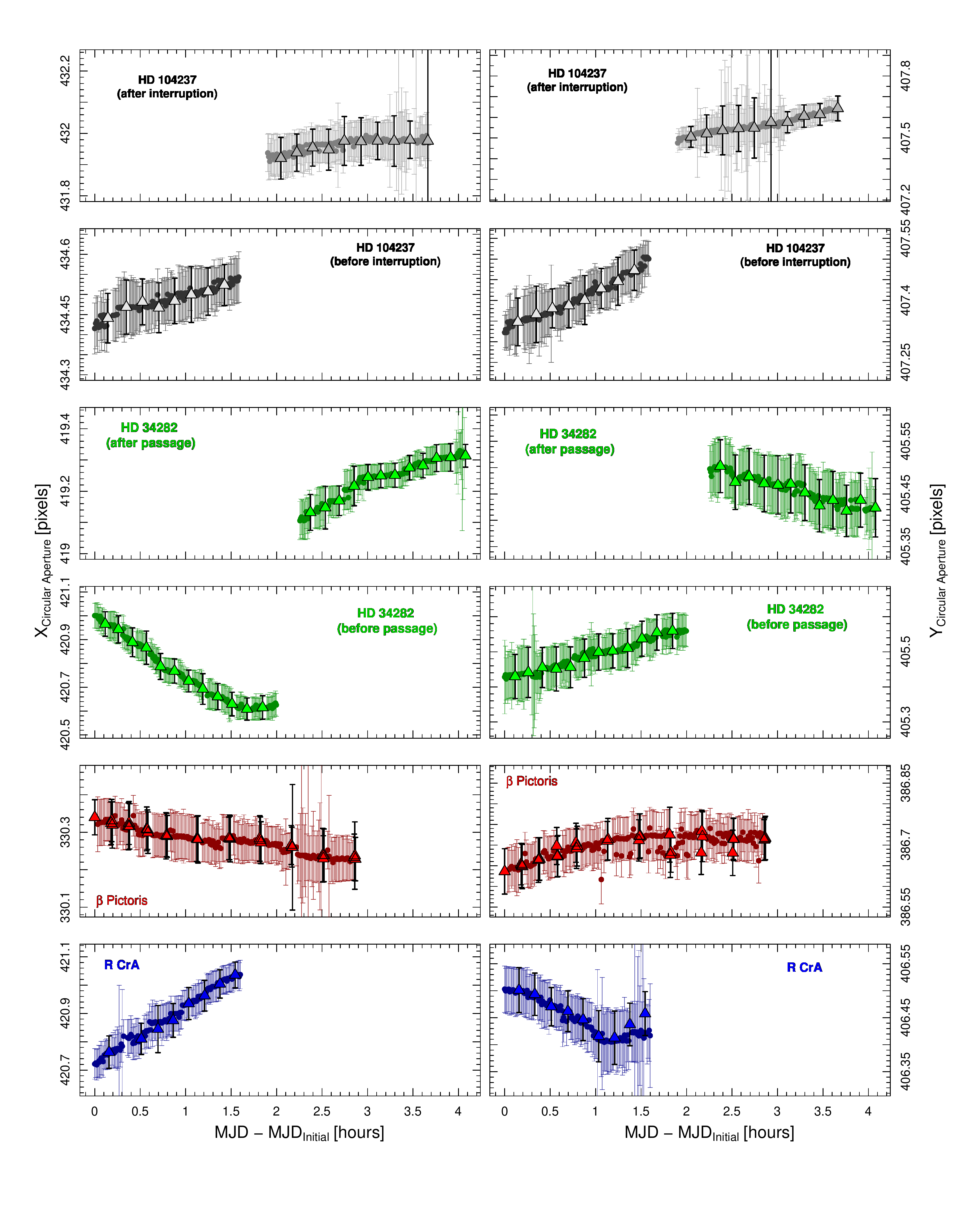}
\caption{\textit{Left panel}: position of the center of the circular aperture along the X-axis as a function of time for all the four targets. If the sequence was interrupted (instrumental problem, or zenith avoidance region), each sequence is plotted in a different panel. The error bars correspond to the 1-$\sigma$ uncertainties. The colored solid circles correspond to science frames, and the colored solid triangles to the sky frames. \textit{Right panel}: same as the left panel but for the Y-axis. }
\label{fig:CA_vs_time}%
\end{figure*}

Figure\,\ref{fig:CA_vs_time} shows the motion of the center of the circular aperture as a function of time, for all four targets. The science frames (colored solid circles), and the sky frames (colored solid triangles) moves in the same random path as a function of time. Even with that random path, the motion of the circular aperture in consecutive science-sky frames is less than $0.1$ pixels.

\section{AGPM - Circular aperture motion as a function of time}\label{Apx:AGPM-CA}

\begin{figure*}
\centering
\includegraphics[width=16cm]{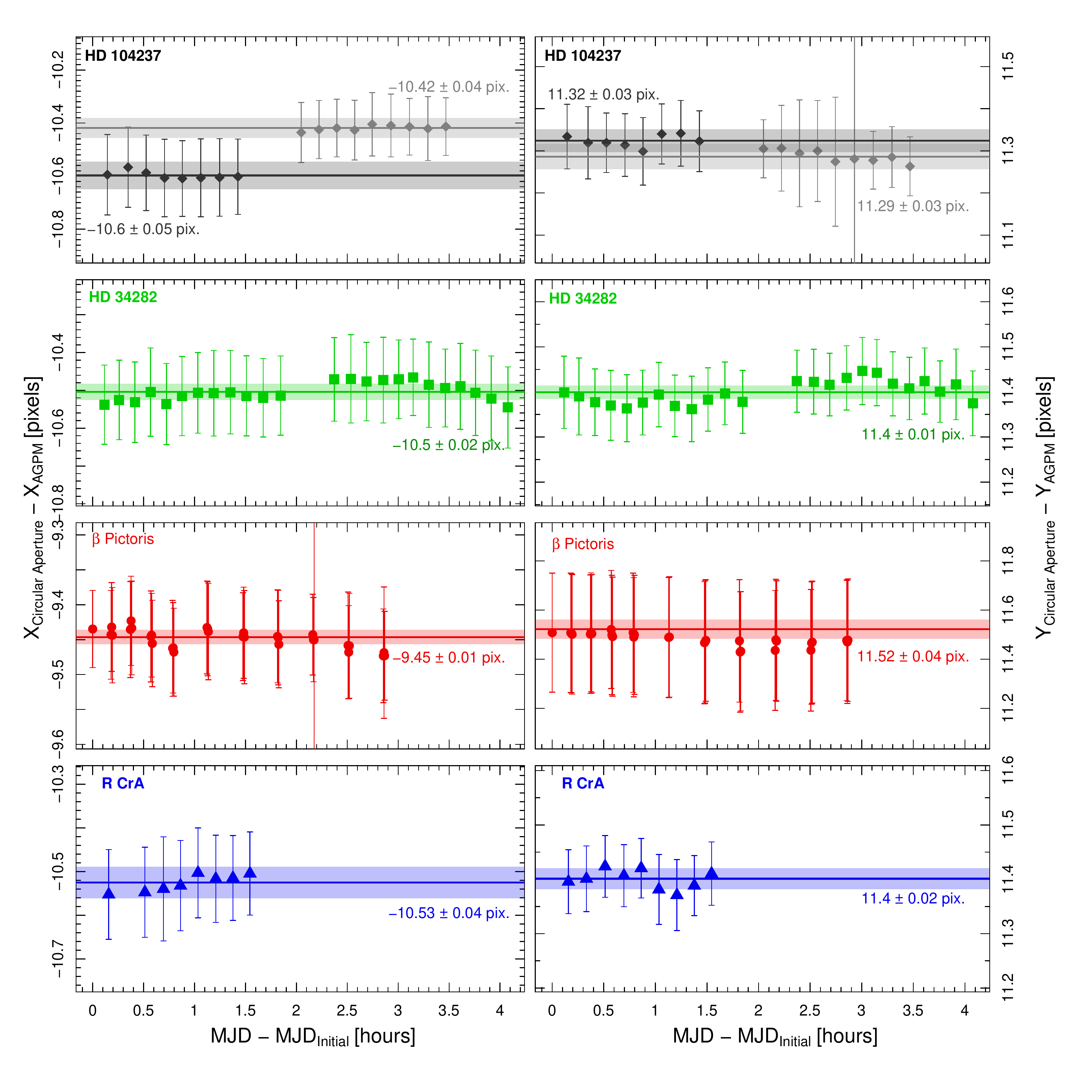}
\caption{Differences between the position of the AGPM and the center of the circular aperture as a function of time (sky frames only), for the X and Y axis (left and right, respectively), for all four targets. The error bars shown represent the $1-\sigma$ uncertainties. The solid horizontal lines correspond to the mean weighted by the uncertainties, and the colored area correspond to the 1-$\sigma$ uncertainty. }
\label{fig:AGPM-CA_vs_time}%
\end{figure*}

Figure\,\ref{fig:AGPM-CA_vs_time} shows the difference between the AGPM position and the circular aperture center as a function of time for both axes X and Y. The relation shows small deviations of higher polynomial order than a constant behaviour, but which are within the data uncertainties at $2-\sigma$ level. To avoid over-fitting, we chose the simplest solution and adopted a constant value calculated from the mean weighted by the respective uncertainties. In addition, Figure\,\ref{fig:AGPM-CA_vs_time} shows the calculated value for every target that provides the direct transformation between the center of the circular aperture and the AGPM position. In the case of HD\,104237 we calculated two different transformations, for each of the observing sequences (before and after the interruption).We further note that as the center of the AGPM is located near the center of the circular aperture, a possible rotation of the circular aperture in time (instead of a linear shift) will not have a strong impact on the differential motion between the circular aperture and the AGPM.

\section{The effect of $H$ in frame selection}\label{Apx:H}

\begin{figure*}
\centering
\includegraphics[width=16cm]{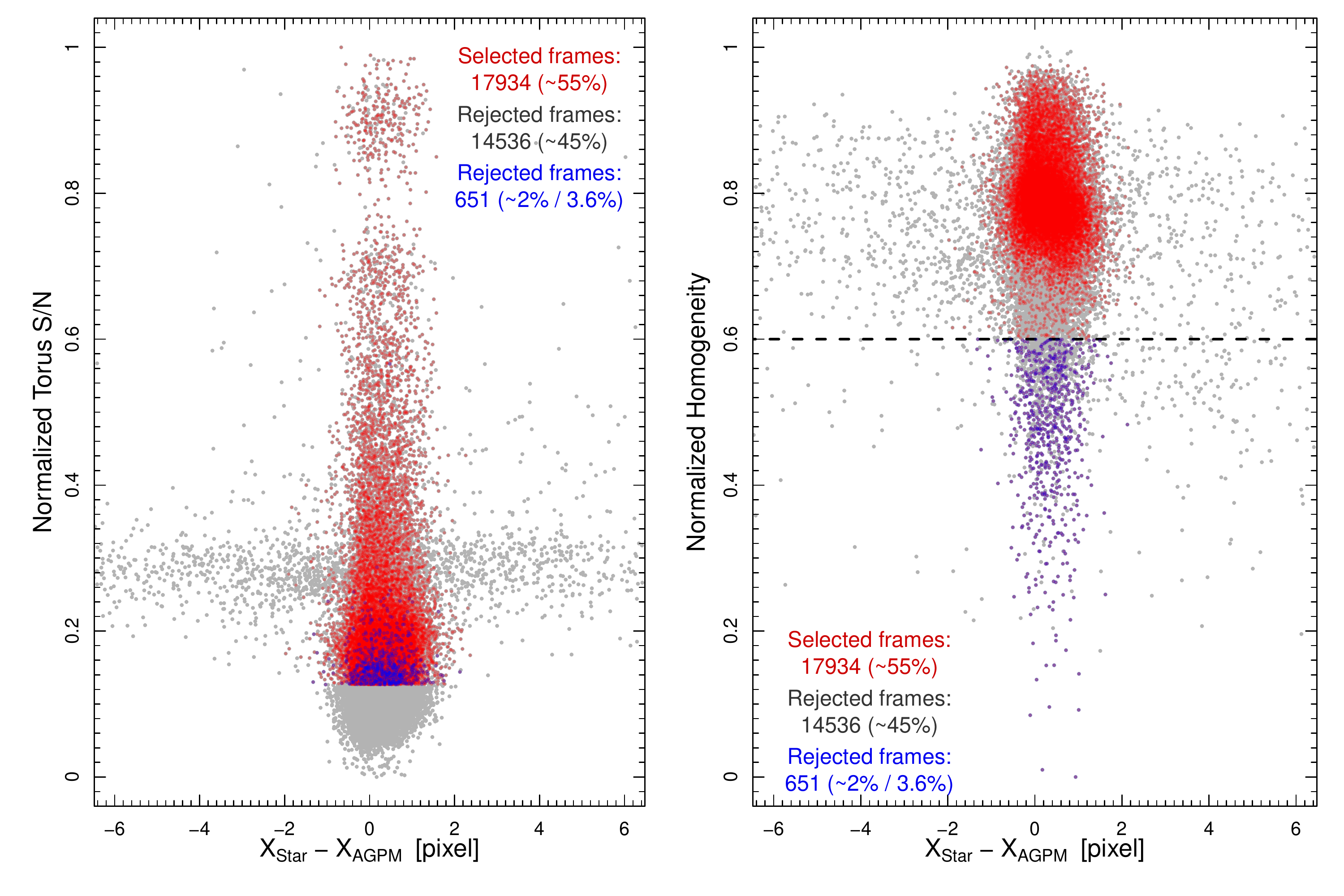}
\caption{\textit{Left:} $\mathrm{Torus_{S/N}}$ as a function of the difference between the star position and AGPM center along the X axis. \textit{Right:} $H$ as a function of the difference between the star position and AGPM center along the X axis. The red dots correspond to the frame selection presented in this work, using $\phi=0.5$ and $\rho=0.5$ corresponding to $\sim55\%$ of the frames, while the grey dots are the rejected frames. The blue dots are the additional rejected frames when $H$ is also considered in the frame selection criteria. The number of blue dots corresponds to $3.6\%$ and $2\%$ of the first selection (red points) and the full dataset, respectively.}
\label{fig:H_vs_Torus}%
\end{figure*}

The homogeneity parameter $H$ is obtained using the reconstruction of pseudo-Zernike moments for the non-radial and azimuthal components (see Eq.\,\ref{eq:Hom_formula} and Fig.\,\ref{fig:Reconstructed_images_psZM_}). The reconstruction is a linear combination where the base corresponds to the pseudo-Zernike moments. For this reason, it is possible to consider the square root sum of those coefficient as a constant (for example, normalizing the root square sum to 1). When the star is well centered behind the AGPM, the torus formed is mostly homogeneous, where the radial component dominates (larger values for $\lvert m \rvert= 0$ coefficients and therefore for $\mathrm{Torus_{S/N}}$, see Eq.\,\ref{eq:SNR_formula} and Fig.\,\ref{fig:Reconstructed_images_psZM_} left panel). When the star is misaligned with respect to the AGPM, the azimuthal components dominate in the reconstruction (larger values for $\lvert m \rvert= 1$ and $2$ coefficients, lower $H$ values, see Eq.\,\ref{eq:Hom_formula} and Fig.\,\ref{fig:Reconstructed_images_psZM_} right panel). Then, $H$ and $\mathrm{Torus_{S/N}}$ are correlated with each other and both provide similar information. Figure\,\ref{fig:H_vs_Torus} shows the normalized $\mathrm{Torus_{S/N}}$ as a function of the star position with respect to the AGPM center along the X-axis in the left panel, and in the right panel the values for $H$ with the same horizontal axis. The red dots correspond to the selected frames ($\sim55\%$), the gray dots to the rejected ones ($\sim45\%$), and the blue dots are the frames that, having been selected in the original frame selection (red dots), are excluded when we also consider the $H$ parameter for the selection, using a cut-off value below $0.6$. We note that both selection criteria are mutually consistent, with a $3.6\%$ discrepancy (corresponding to the $\sim2\%$ of all the frames). The location of these blue dots in both graphics shows that both $\mathrm{Torus_{S/N}}$ and $H$ classify those frames as low quality frames with respect to the other ones. Therefore, the incorporation of $H$ parameter does not have a strong impact on data processing, since it provides information similar and consistent with $\mathrm{Torus_{S/N}}$.

\section{Relative positions and Torus S/N}\label{Apx:Rel_Pos_Torus}

\begin{figure*}
\centering
\includegraphics[width=16cm]{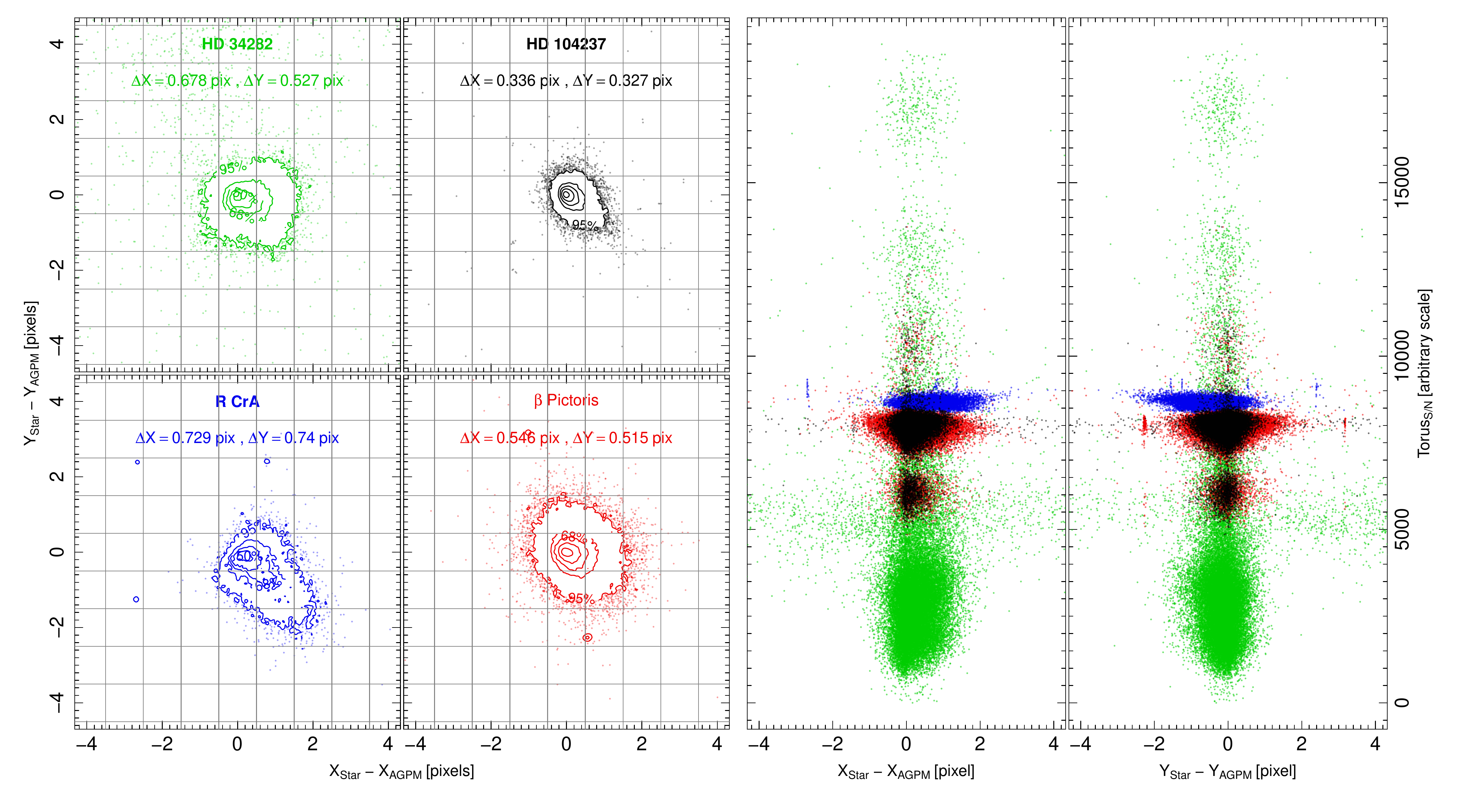}
\caption{\textit{Left panel}: distribution of the difference between the positions of the star and of the AGPM for all four stars included in this study. The $\Delta$X and $\Delta$Y correspond to $\sim 1.5$ times the median absolute deviation. \textit{Right panel}: the Torus$_\mathrm{S/N}$ for all the targets in a common reference frame. The contours contain the $10\%$, $30\%$, $50\%$, $68\%$ and $95\%$ of the data, respectively.}
\label{fig:Rel_Pos_and_Torus_S2N_unnorm}%
\end{figure*}

The left panel of Fig.\,\ref{fig:Rel_Pos_and_Torus_S2N_unnorm} shows the position of the stars with respect to the AGPM for all the sources. The right panel shows the de-normalized Torus$_\mathrm{S/N}$ for a more direct comparison between all the observations at different weather conditions. In particular, R\,CrA shows a not centro-symmetric distribution. This implies that the choice of the center becomes more critical, affecting the S/N and the astrometric measurements of the companion. In contrast, $\beta$\,Pictoris shows a more uniform distribution (azimuthally homogeneous), where we do not observe significant differences in the results for the S/N and positions of the companion.

\section{Frames of R CrA}\label{Apx:frames_corona}

Regions of the frames of R\,CrA are saturated, especially close to the AGPM in the entire observing sequence, resulting in stronger signal of the torus and speckles around the AGPM. Figure\,\ref{fig:RCrA_disalig_fig} shows the distribution of the star position with respect to the center of the AGPM along both axes (top-left panel) and with respect to the $\mathrm{Torus_{S/N}}$ and homogeneity $H$ (top-right and bottom-left panel, respectively). Examples of the different images at different separations are shown in the bottom-right panel, corresponding to examples from good to poor centering (red triangles, green squares and cyan circles, respectively). In this particular case, both $\mathrm{Torus_{S/N}}$ and $H$ are less informative to distinguish between frames with good and poor centering. However, the distribution of the star position with respect to the AGPM is a more robust criterion to select the most homogeneous and well-centered frames.

\begin{figure*}
\centering
\includegraphics[width=14.5cm]{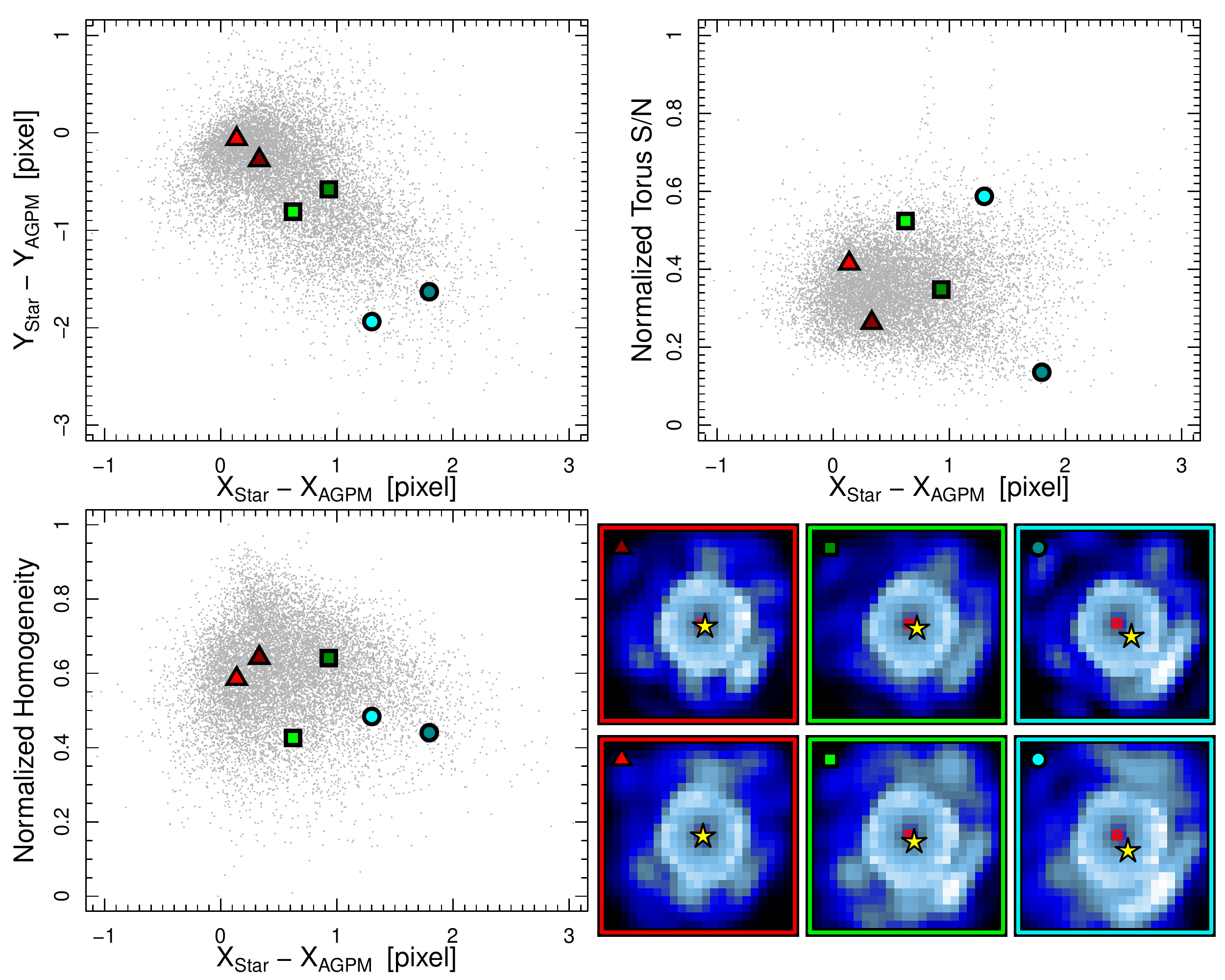}
\caption{\textit{Top left}: distribution of the star position with respect to the AGPM center along both axes. \textit{Top right}: $\mathrm{Torus_{S/N}}$ as a function of the star position with respect to the AGPM center along the X-axis. \textit{Bottom left}: homogeneity $H$ as a function of the star position with respect to the AGPM center along the X-axis \textit{Bottom right}: examples of images with good (red triangles), intermediate (green square), and poor (cyan circles) centering. In the images, the yellow stars correspond to the star position, while the red crossed circles mark the AGPM position. The images have the same color scale.}
\end{figure*}\label{fig:RCrA_disalig_fig}%

\section{Position of companions as a function of number of components}\label{Apx:Comp_vs_nPC}

We made a 2D Gaussian fitting to each of the companions ($\beta$\,Pictoris\,b and R\,CrA\,b) to know their approximate positions in each of the post-processed images according to the number of components used. Since this method of knowing the position of the object in question is very sensitive to residual speckles and then to the number of components used, we took all the images to calculate the weighted average by the uncertainties associated with each position. Figure\,\ref{fig:BP_pos_vs_nPC} shows an example for $\beta$\,Pictoris\,b of the results of our fitting for the X and Y axis (solid points with their error bars), as well as the calculated average position and its associated uncertainty (horizontal solid line and the colored area, respectively), for the case of full data cubes. Figure\,\ref{fig:RC_pos_vs_nPC} shows an example for R\,CrA\,b in the same format. These figures show the trend of the positions of each of the objects as a function of the number of components. It is noticeable that, as the number of components increases, the associated uncertainty also increases due to the fact that there is less flux available for the fitting (effect of self-subtraction).

\begin{figure*}
\centering
\includegraphics[width=16cm]{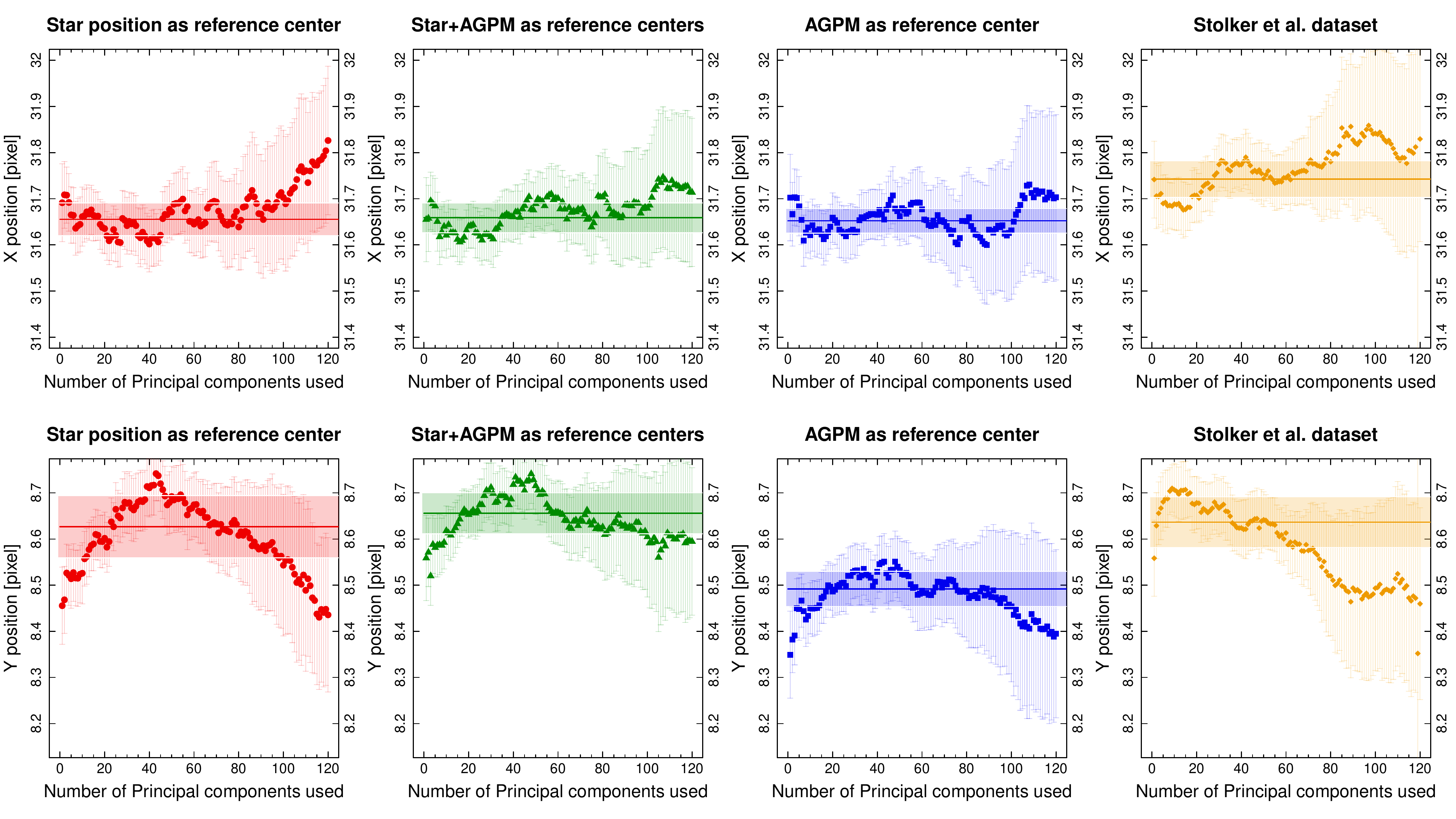}
\caption{Each panel shows the position for X (top panels) and Y (bottom panels) using our 2D Gaussian fitting for $\beta$\,Pictoris\,b. From left to right, the center used for our reduced data correspond to: only the star, the star and the AGPM, only the AGPM, and at the end the dataset from \cite{Stolker+2019}. The solid horizontal line correspond to the weighted mean by the respective uncertainties (the vertical lines in each solid point), with the respective associated uncertainty (colored horizontal area).}
\end{figure*}\label{fig:BP_pos_vs_nPC}%

\begin{figure*}
\centering
\includegraphics[width=15cm]{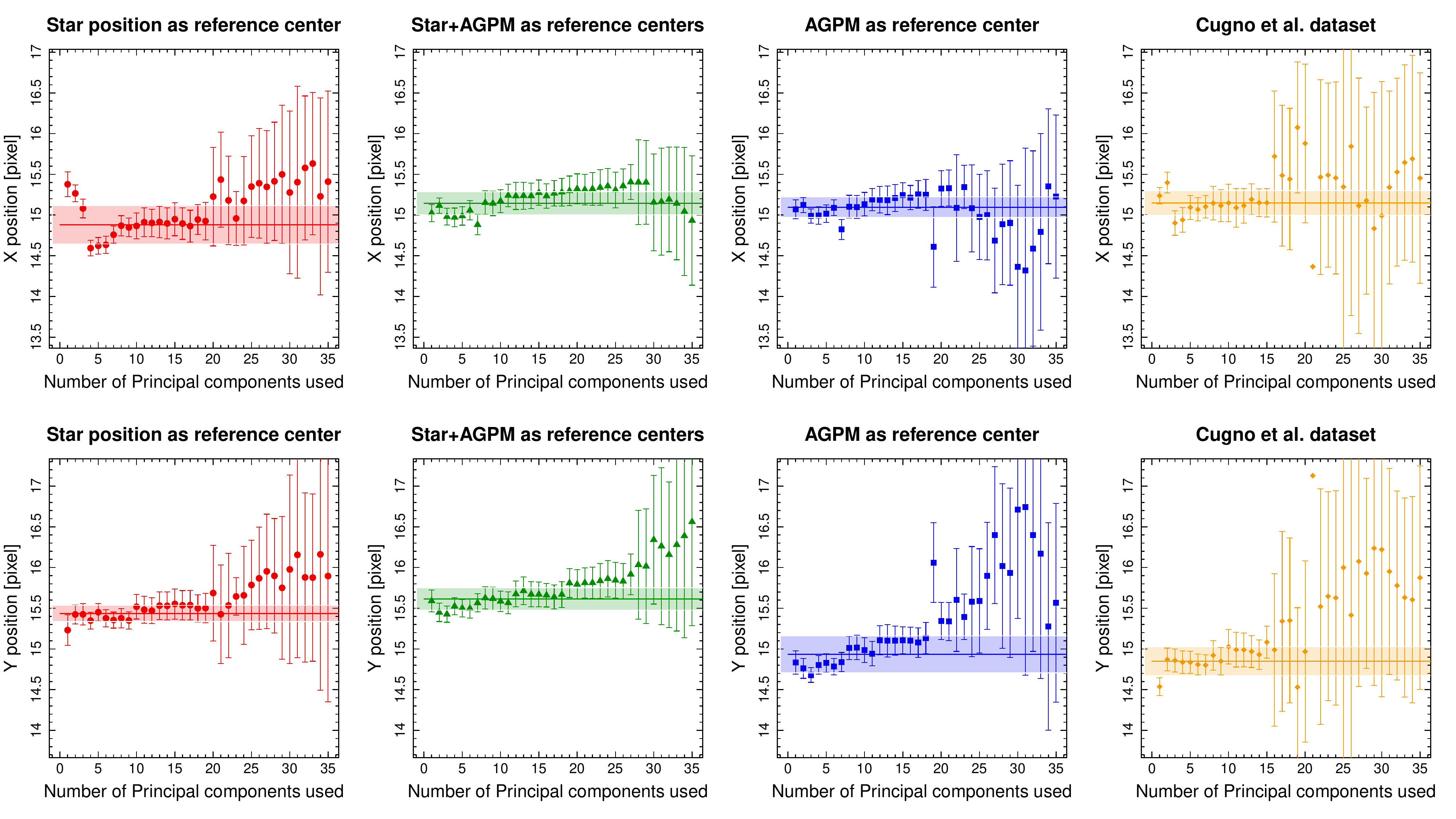}
\caption{Same as Fig.\,\ref{fig:BP_pos_vs_nPC} but for R\,CrA\,b. }
\end{figure*}\label{fig:RC_pos_vs_nPC}%

\section{Extended sources}\label{Apx:Disk}

The advantage of using direct imaging, compared with polarimetric observations, is that it is possible to obtain the full intensity of extended structures. However, it is usually a challenge for angular differential imaging as it removes part of the flux and can introduce dark regions at the edges. In most cases, the circumstellar disks are significantly removed in the final image especially for disk with inclinations smaller than $60^{\circ}$ (see \citealt{Milli+2012}), as a disk with low inclination will be incorporated in the reference PSF to be subtracted. Different efforts have been put into addressing and solving this problem, either from the observations themselves (for example, RDI, \citealt{Smith_and_Terrile_84}), or in post-processing techniques (see, for instance, LOCI, \citealt{Lafreniere+2007}). More recently, \citet{Ren+2018,Ren+2020} implement a new algorithm, the data imputation using sequential non-negative matrix factorization (DI-sNMF), that is a model-free solution in a post-detection disk scenario. Broadly speaking, the idea is to transform the signal (disk or point-sources) into a missing data problem, assigning to that region the PSF signal (following the statistical properties). Then, it should be possible to remove both the stellar PSF and the speckles in every image, avoiding problems of self-subtraction. However, none of these studies investigate the effect of the centering or frame selection on the properties of the resulting disk image (signal, structure, self-subtraction), and detectability (noise distribution, significance of the detection, for instances).     

We here study the impact of both the centering and frame selection when the target shows extended emission in its vicinity. From the previous sections, we know that the centering has a significant impact on the results for point sources, but the effect of frame selection is marginal. However, in the case of disks, frame selection may improve the speckle suppression. If we keep the most homogeneous, well-centered frames, we should be able to decrease the number of components needed, overall decreasing the contribution of residual speckles in the inner regions. Furthermore, removing the frames with a poor quality (poor centering or weather conditions), the contrast should improve in the inner regions where the disk is located. We tested our reduction strategy on the HAeBe star HD\,34282. The star is located at a distance of $359\pm5$\,pc \citep{Gaia_DR2}, with an age of $6.41^{+1.92}_{-2.58}$\,Myr \citep{Merin+2004}. The disk was first resolved with ALMA observations \citep{van_der_Plas+17}, and first imaged at infrared wavelength by \citet{ISPY+2020} as part of the ISPY program using VLT/NaCo, and by \citet{deBoer_2021} as part of the DISK program in polarimetric mode at near-infrared using VLT/SPHERE. The reduction is performed in the same way as for the previous targets of this paper. The details related to the observations are summarized in Table\,\ref{Summary_table}. We performed the frame selection using $\phi$ and $\rho$ as $0.5$, keeping $\sim50$\,\% of $32\,825$ available frames.

Figure\,\ref{fig:HD_images_disk} shows reduction examples for three different centering configurations (top to bottom), using $18$ principal components. From left to right, we show the final images when using the full dataset, with our frame selection, and a ``control'' reduction, selecting at random the same number of frames as for the middle column. Quantifying the S/N for extended emission is more difficult than for a point source, therefore, we just compare the images visually, paying special attention to the noise level as well as self-subtraction effects. For instance, when using the full dataset, the dark pattern caused by self-subtraction is quite significant, alongside with spots toward the east. In contrast, the reduction using frame selection shows a better removal of several speckles, consequently the structure of the disk is overall better recovered. This can be explained by the presence of a significant number of poorly centered frames in the entire dataset, which increases the noise in the innermost regions and dilutes the faint signal from the disk.  

When performing frame selection, we may remove a significant number of frames, and we therefore introduce gaps in the sequence of frames. Those gaps correlate with the observing condition or the AO performance. As a consequence, we most likely remove series of consecutive frames, which may act as a ``protection angle''\footnote{The ``protection angle'' consists of removing some of the adjacent frames in the time series. For the $i^{\mathrm{th}}$-frame in the sequence, only the frames for which the paralactic rotation is larger than $\Delta\theta$ are used to build the reference, where $\Delta\theta$ depends on the angular size of the PSF at a specific separation from the center. The selected frames are used in the ADI, and this approach minimizes the effect of self-subtraction.} (see LOCI, \citealt{Lafreniere+2007}), therefore decreasing the effect of self-subtraction. To understand if the result presented in Figure\,\ref{fig:HD_images_disk} (middle panel) is due to the frame selection or the result of an non-homogeneous sampling in time, we carry out the following test. First, from the cube with frame selection (using $\phi=0.5$ and $\rho=0.5$), we count the number and sizes of the gaps introduced in the sequence. We then clone this distribution to obtain a new sample of gaps, which we can randomly introduce in the entire dataset. This new sub-sample has the same number of frames as the previous one. The result is shown in Figure\,\ref{fig:HD_images_disk} (right column). Overall, the signal from the disk appears fainter, the western side seems to be more disconnected from the trace of the disk, and the speckles on the eastern side are still visible while they had disappeared when using our frame selection approach. This result strongly suggests that our strategy to perform frame selection does improve the final reduction, and that it is not solely related to the sampling of frames as a function of time. We also note that, in agreement with the previous section, using only the AGPM position as the reference center does not yield to the best final image. Without a proper quantification of the S/N of the disk, it remains difficult to estimate which of the other centering strategies provides the best reduction.

\begin{figure*}
\centering
\includegraphics[width=15cm]{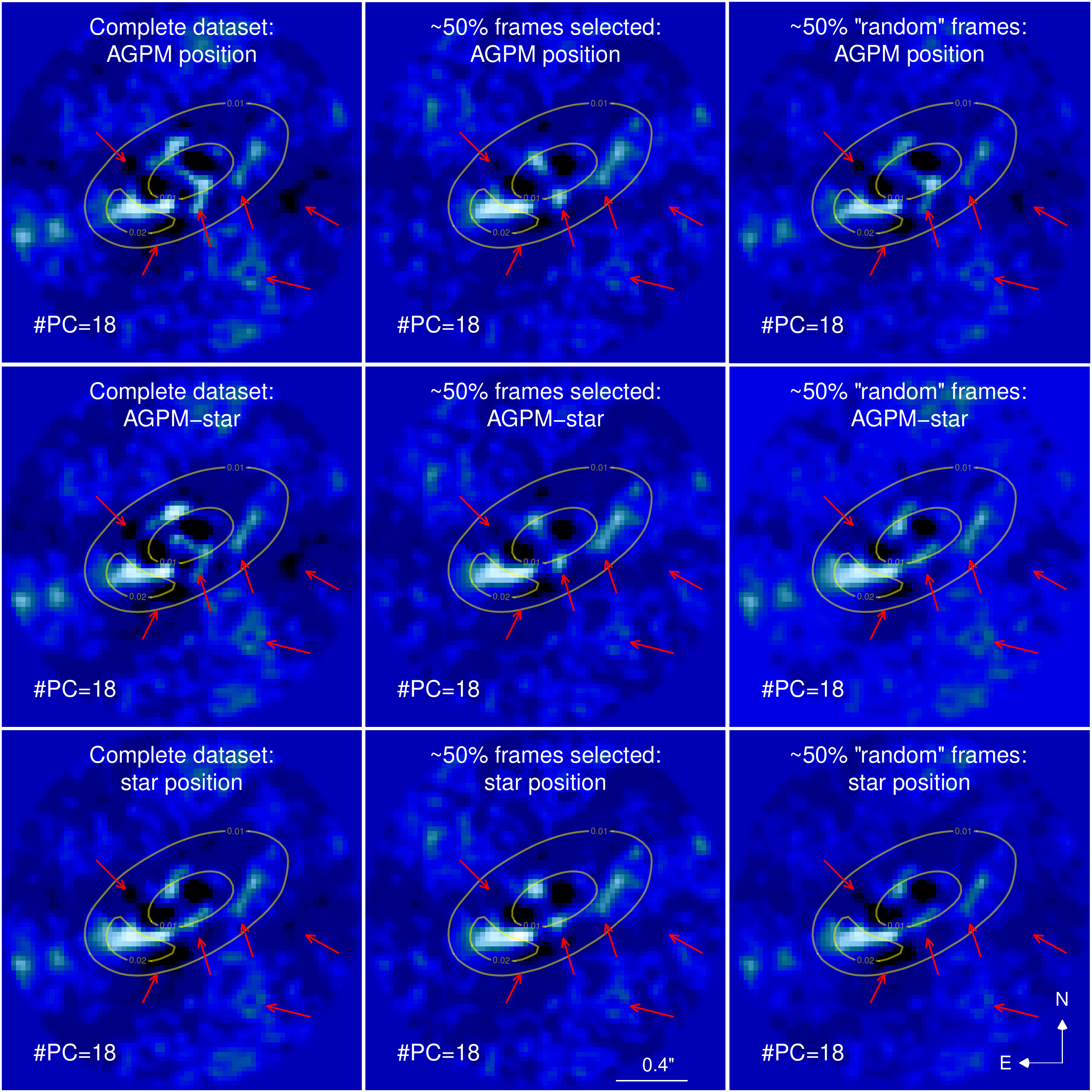}
\caption{\textit{Left panel}: final images of HD\,34282 using the entire dataset but implementing three different centers, that is, using the AGPM position only, both the AGPM and the star positions, and using only the star position. \textit{Middle panel}: same as the previous panel but with frame selection, using $\sim50$\% of the data. \textit{Right panel}: same as the middle panel but with a ``random'' selection of frames (see text for details). Red arrows mark the most noticeable lingering shadows and speckles, as well as the most notable disk features. The yellow contours correspond to the continuous image of the observations made with ALMA (\citealt{van_der_Plas+17}), and correspond to 10 and 20 mJy levels.}
\end{figure*}\label{fig:HD_images_disk}%

\end{appendix}

\end{document}